\documentclass[journal]{IEEEtran}
\usepackage{graphicx}
\usepackage{comment}
\usepackage{amsmath,amssymb} % define this before the line numbering.
\usepackage{color}
\usepackage{overpic}
\usepackage{multirow}
\usepackage{subfigure}
\usepackage{tabularx}
\usepackage{adjustbox}
\usepackage{booktabs}
\usepackage{setspace}
\newcolumntype{P}[1]{>{\centering\arraybackslash}p{#1}}
\begin{document}
\title{Joint Denoising and Demosaicking with \\ Green Channel Prior for Real-world Burst Images}
\author{Shi~Guo,
        Zhetong~Liang,
        and~Lei~Zhang,~\IEEEmembership{Fellow,~IEEE}% <-this % stops a space
\thanks{S. Guo, Z. Liang and L. Zhang are with the Hong Kong Polytechnic University, Kowloon, Hong
Kong and DAMO Academy, Alibaba Group. E-mail: \{csshiguo, csztliang, cslzhang\}@comp.polyu.edu.hk.}% <-this % stops a space
}
\maketitle

\begin{abstract}
Denoising and demosaicking are essential yet correlated steps to reconstruct a full color image from the raw color filter array (CFA) data. By learning a deep convolutional neural network (CNN), significant progress has been achieved to perform denoising and demosaicking jointly. However, most existing CNN-based joint denoising and demosaicking (JDD) methods work on a single image while assuming additive white Gaussian noise, which limits their performance on real-world applications. In this work, we study the JDD problem for real-world burst images, namely JDD-B. Considering the fact that the green channel has twice the sampling rate and better quality than the red and blue channels in CFA raw data, we propose to use this green channel prior (GCP) to build a GCP-Net for the JDD-B task. In GCP-Net, the GCP features extracted from green channels are utilized to guide the feature extraction and feature upsampling of the whole image. To compensate for the shift between frames, the offset is also estimated from GCP features to reduce the impact of noise. Our GCP-Net can preserve more image structures and details than other JDD methods while removing noise. Experiments on synthetic and real-world noisy images demonstrate the effectiveness of GCP-Net quantitatively and qualitatively.
\end{abstract}


% \fi
\IEEEpeerreviewmaketitle

\section{Introduction}
\IEEEPARstart{M}{ost} consumer grade digital cameras capture natural images using a single-chip CCD/CMOS sensor covered by a color filter array (CFA), resulting in incomplete color sampling at each photoreceptor. The process of interpolating the missing colors from mosaicked CFA data is called color demosaicking. The captured data is inevitably corrupted by noise, especially under low-light conditions. Denoising and demosaicking play crucial roles to obtain high quality images in the camera ISP (image signal processing) pipeline, and a variety of image denoising and demosaicking methods~\cite{ehret2019study,yang2019efficient,yan2019cross,liu2020new,jin2020review} have been proposed.

Previous demosaicking and denoising methods are usually designed independently and implemented sequentially in the ISP. However, the demosaicking errors will complicate the denoising process, or the denoising artifacts can be amplified in the demosaicking process. Therefore, joint denoising and demosaicking (JDD) has received considerable research interests~\cite{condat2012joint,heide2014flexisp,tan2017joint,gharbi2016deep,henz2018deep,ehret2019joint,kokkinos2019iterative,qian2019trinity}. Traditional JDD methods resort to image priors, such as piecewise smoothness~\cite{condat2012joint} and non-local self-similarity~\cite{heide2014flexisp}, and employ an optimization model for this joint task. Those handcrafted priors, however, are not accurate enough to reproduce the complex image local structures. Recent JDD methods are mostly data-driven learning methods, where a deep convolutional neural network (CNN) is trained on pairwise dataset with noisy mosaicked images and their clean full color ground truths~\cite{gharbi2016deep,henz2018deep,kokkinos2019iterative,qian2019trinity,liu2020joint}. By learning deep priors from a large amount of data, those CNN based methods achieve much better JDD performance than traditional model based methods.

\begin{figure}[!t]
\setlength{\abovecaptionskip}{0.2cm}
%\setlength{\belowcaptionskip}{-0.cm}
%\centering
\begin{minipage}[b]{0.48\textwidth}
\centering
    \begin{minipage}[b]{1\textwidth}
    \centering
    \subfigure{
        \includegraphics[width=1\textwidth]{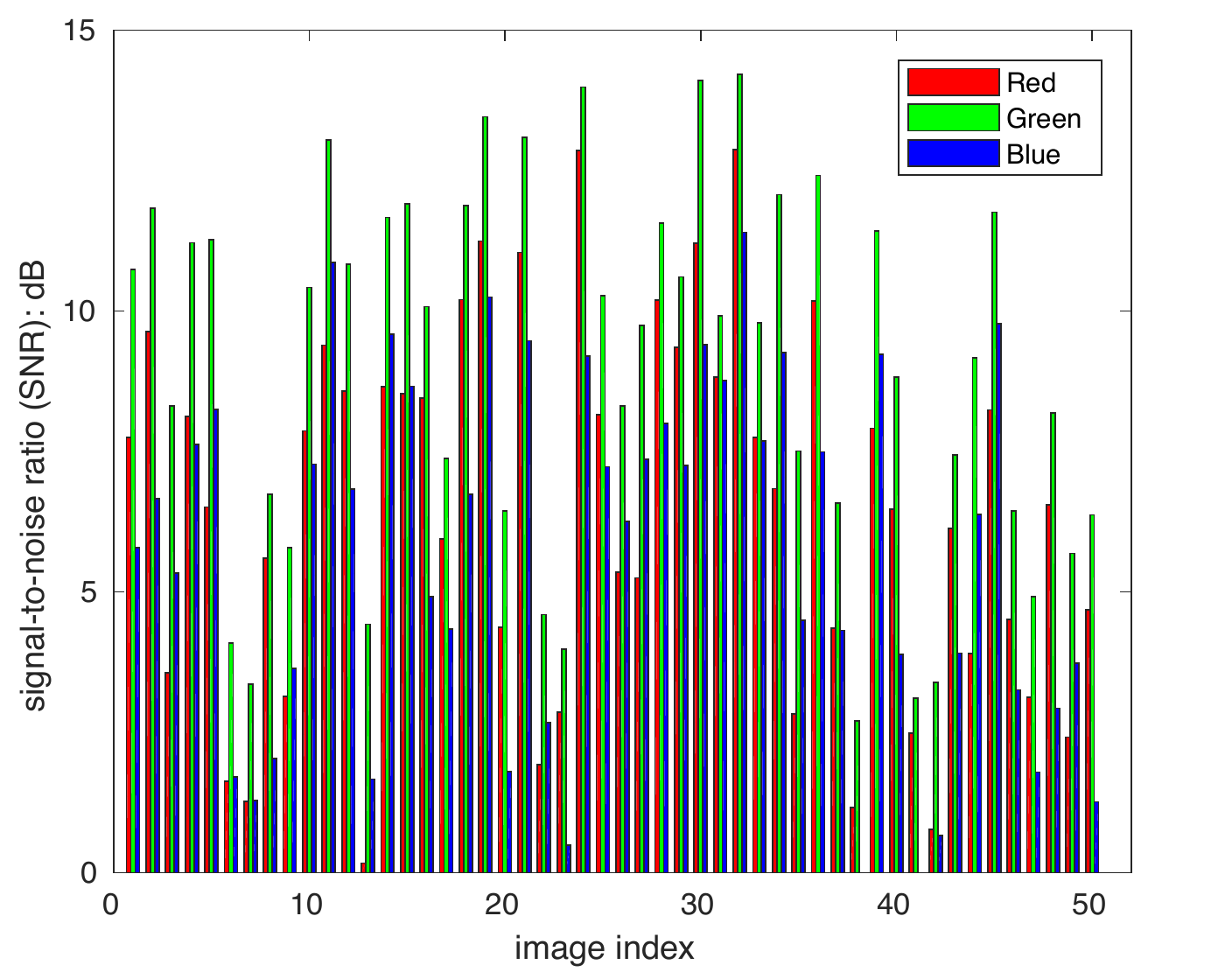}
    }
    % {\footnotesize  (a)}
    \end{minipage}
\end{minipage}\hspace{-2ex}

\caption{Signal-to-noise ratio (SNR) in different color channels ({\color{red} Red}, {\color{green} Green} and {\color{blue} Blue}) of 50 high-ISO noisy images randomly chosen from the SIDD dataset~\cite{abdelhamed2018high}. Clearly, the green channel has the best SNR.}%\vspace{-0.4cm}
\label{fig:noise_snr}
\end{figure}

Existing JDD methods, including those CNN based ones, mostly work on a single CFA image, and we call them JDD-S methods, which have several limitations when applying to real-world CFA data. First, their performance will deteriorate significantly on CFA images with strong noise level. This often occurs for low-end devices such as smartphone cameras due to the small sensor and lens. The situation becomes worse under low-light imaging conditions. Second, current JDD-S methods~\cite{gharbi2016deep,henz2018deep,kokkinos2019iterative,qian2019trinity,liu2020joint} usually assume additive white Gaussian noise (AWGN) in the training process, which cannot accurately describe the distribution of real-world noise. As a result, strong visual artifacts will appear in the JDD outputs of real-world noisy CFA images. 

Recently, it has been shown that the denoising performance can be significantly improved by using a set of burst images instead of a single image, especially for the low-light imaging conditions~\cite{mildenhall2018burst,xue2019video,ehret2019model,wang2019edvr,yue2020supervised}. Inspired by the success of burst image denoising, we propose to perform JDD with real-world burst images, which is called JDD-B. With some realistic noise modeling methods~\cite{mildenhall2018burst,brooks2019unprocessing}, we can synthesize noisy burst images with clean ground truth by reversing the ISP pipeline on high quality video sequences and adding noise into them. Such pairwise data can be used to train the JDD-B model. It is well-known that the green channel of images captured by single-chip digital cameras has better quality than red and blue channels. On one hand, the green channel has twice the sampling rate than red/blue channels in most CFA patterns (\emph{e.g.}, Bayer pattern). On the other hand, the sensitivity of green is better than red/blue~\cite{tan2014green}. As a result, the green channel has more texture information and higher SNR than red/blue channels in most natural images, which is demonstrated in Fig.~\ref{fig:noise_snr} by using the SIDD dataset~\cite{abdelhamed2018high}. We call the above fact and prior knowledge the green channel prior (GCP), and use the GCP to design our JDD-B network, namely GCP-Net, to improve the JDD-B performance on real-world burst images.

Specifically, in GCP-Net, we extract the GCP features from green channel to guide the deep feature modeling and upsampling of the whole image. The GCP features are also utilized to estimate the offset within frames to relief the impact of noise. As shown in Fig.~\ref{figcover}, with GCP, the JDD-B results can preserve more structures and details while removing noise. Our GCP-Net achieves state-of-the-art JDD-B performance on both synthetic noisy images and real-world burst images captured by smartphones.

The remaining of this paper is organized as follows. Sec.~\ref{related} reviews existing methods for JDD and burst image restoration. Sec.~\ref{method} states the problem of JDD-B and presents the proposed GCP-Net in detail. Sec.~\ref{exp} reports the ablation study and experimental results. Sec.~\ref{con} concludes the paper.

\section{Related Work}
\label{related}
\subsection{Joint Denoising and Demosaicking for Single Image}
Image denoising and demosaicking are two important steps in camera ISP pipeline. A few methods have been proposed for joint denoising and demosaicking on a single raw image (JDD-S)~\cite{gharbi2016deep,tan2017color,henz2018deep,ehret2019joint,kokkinos2019iterative,qian2019trinity,liu2020joint}. In~\cite{qian2019trinity}, Qian \emph{et al.} showed that the performance of JDD-S is generally better than performing denoising and demosaicking separately. In \cite{gharbi2016deep}, a learning-based method was proposed for JDD-S. Henz \emph{et al.}~\cite{henz2018deep} proposed an auto-encoder architecture to model the color-image capturing process on each monochromatic sensor. Kokkinos \emph{et al.}~\cite{kokkinos2019iterative} proposed a plug-and-play framework for the JDD-S task. To enhance the performance on real-world images, Ehret \emph{et al.}~\cite{ehret2019joint} proposed a mosaic-to-mosaic framework by finetuning the network using mosaic burst images. It should be noted that though Ehret \emph{et al.} utilized burst images to fine-tune the network, the input of the network is still a single image. Liu \emph{et al.}~\cite{liu2020joint} proposed a self-guided JDD-S network by considering the advantages of the higher sampling rate of green channel and using this prior to guide the upsampling process. In this paper, we further analyze the noise level imbalance among different color channels in real-world photographs, and perform the JDD task using burst images instead of a single image.

\begin{figure}[!h]
\setlength{\abovecaptionskip}{0.2cm}
%\setlength{\belowcaptionskip}{-0.cm}
%\centering
\begin{minipage}[b]{0.3375\textwidth}
\centering
    \begin{minipage}[b]{1.0\textwidth}
    \centering
    \subfigure[hang,scriptsize,tight][Noisy Image]{
        \includegraphics[trim=0 0 0 5.0cm, clip, width=1\textwidth]{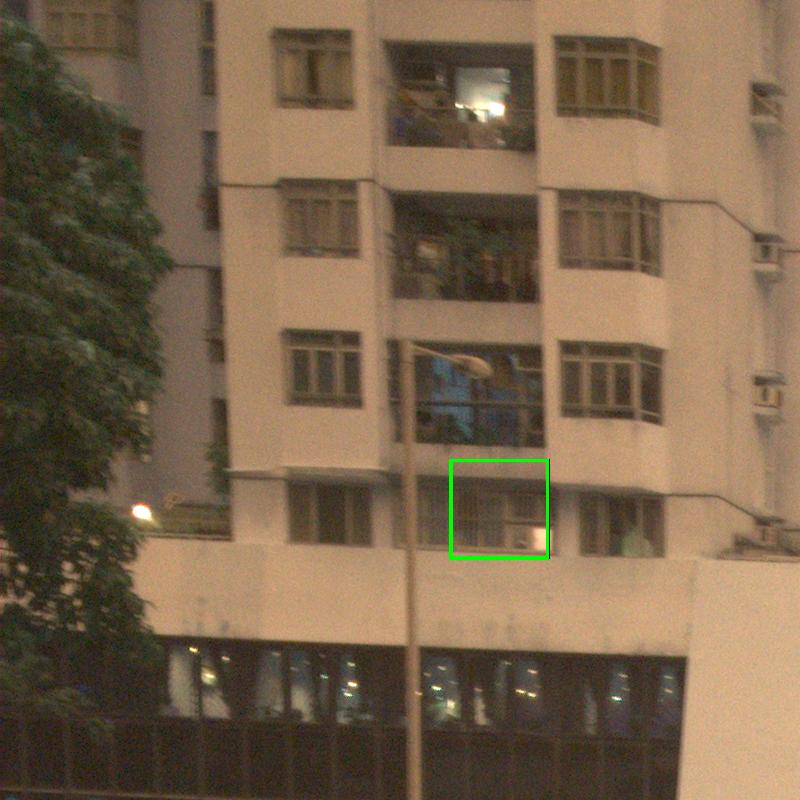}
        }
    % {\footnotesize  (b)}
    \end{minipage}
    %\vspace{0.10cm}
\end{minipage} \hspace{-1ex}
\begin{minipage}[b]{0.13\textwidth}
    \begin{minipage}[b]{1\textwidth}
    \centering
        \begin{minipage}[b]{1\textwidth}
        \centering
        \subfigure[hang,scriptsize,tight][w/o GCP]{
            \includegraphics[width=1\textwidth]{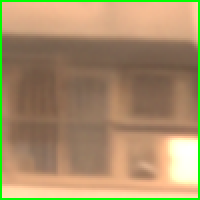}
        }
        % {\footnotesize  w/o GG}
        \end{minipage} 

        \begin{minipage}[b]{1\textwidth}
        \centering
        \subfigure[hang,scriptsize,tight][Full model]{
            \includegraphics[width=1\textwidth]{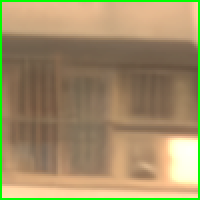}
        }
        % {\footnotesize  Full model}
        \end{minipage} 
    \end{minipage}%\hspace{0.1cm}
    %\vspace{0.15cm}
\end{minipage}\hspace{-2ex}

\begin{minipage}[b]{0.48\textwidth}
\hspace{-1ex}
%\centering
    \begin{minipage}[b]{1\textwidth}
    \centering
        \begin{minipage}[b]{0.185\textwidth}
        \centering
        \subfigure[hang,scriptsize,tight][R]{
            \includegraphics[width=1\textwidth]{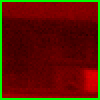}
        }
        %{\footnotesize  R}
        \end{minipage} %\hspace{0.1cm}
        \begin{minipage}[b]{0.185\textwidth}
        \centering
        \subfigure[hang,scriptsize,tight][Gr]{
            \includegraphics[width=1\textwidth]{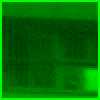}
        }
        %{\footnotesize Gr}
        \end{minipage} %\hspace{0.1cm}
        \begin{minipage}[b]{0.185\textwidth}
        \centering
        \subfigure[hang,scriptsize,tight][Gb]{
            \includegraphics[width=1\textwidth]{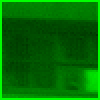}
        }
        %{\footnotesize  Gb}
        \end{minipage} 
        \begin{minipage}[b]{0.185\textwidth}
        \centering
        \subfigure[hang,scriptsize,tight][B]{
            \includegraphics[width=1\textwidth]{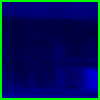}
        }
        %{\footnotesize  B}
        \end{minipage} 
        \begin{minipage}[b]{0.185\textwidth}
        \centering
        \subfigure[hang,scriptsize,tight][RGB]{
            \includegraphics[width=1\textwidth]{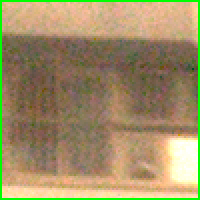}
        }
        %{\footnotesize  noise}
        \end{minipage} %\hspace{0.1cm}
    \end{minipage}%\hspace{0.1cm}
    %\vspace{0.15cm}
\end{minipage}\hspace{-2ex}
\caption{(a) shows a noisy image captured using iPhoneX and (h) shows a patch of it, while (d) - (g) show the noisy patches of different color channels. Gr and Gb mean two green channels. (b) and (c) are the JDD results of our network without and with the green channel prior (GCP).}%\vspace{-0.4cm}
\label{figcover}
\end{figure}

\begin{figure*}[!t]
\centering
%\hspace{-0.6cm}
\begin{overpic}[width=1.0\textwidth]{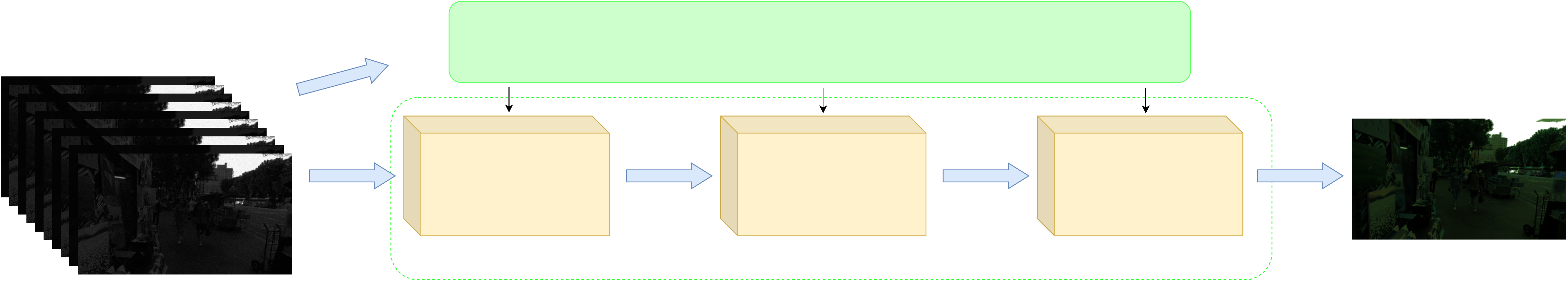}
    \put(18,15){\color{black}{\footnotesize $\mathcal{Y}_g, \mathcal{M}_g$}}
	\put(20,8){\color{black}{\footnotesize $\mathcal{Y}, \mathcal{M}$}}
	\put(28,5.5){\color{black}{\footnotesize \textbf{IntraF Module}}}
	%\put(31.0,4.0){\color{black}{\footnotesize $F_{\text{GPF}}$}}
	\put(39.5,8){\color{black}{\footnotesize $\{ f_t \} _{t=1}^N$}}
	
	\put(48,5.5){\color{black}{\footnotesize \textbf{InterF Module}}}
	%\put(51.5,4.0){\color{black}{\footnotesize $F_{\text{IF}}$}}
	\put(59.5,8){\color{black}{\footnotesize $\{ f^a_t \} _{t=1}^N$}}
	
	\put(68.5,5.5){\color{black}{\footnotesize \textbf{Merge Module}}}
	%\put(71.5,4.0){\color{black}{\footnotesize $F_{\text{M}}$}}
	\put(81,8){\color{black}{\footnotesize $\hat{x}_{ref}^{rgb}$}}
	
	\put(40,14.5){\color{black}{\footnotesize \textbf{Green Channel Prior (GCP) Branch}}}
	\put(45,1.2){\color{black}{\footnotesize \textbf{Reconstruction Branch}}}
	
\end{overpic}

\caption{Overview of our GCP-Net for JDD-B.}
\label{figFlow}
%\vspace{-0.3cm}
\end{figure*}
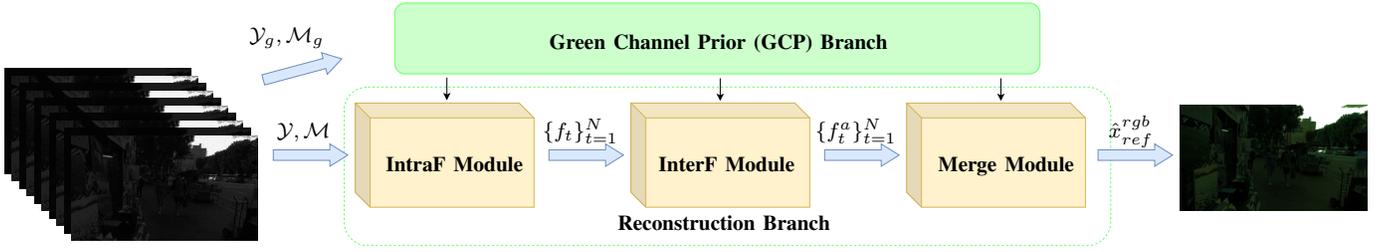

\subsection{Burst Image Restoration}
Compared with single image restoration tasks, burst image processing encounters new challenges on estimating the offsets among different frames caused by camera movement and moving objects. According to the employed alignment frameworks, we partition burst image restoration methods into three categories, \emph{i.e.}, pre-aligned methods~\cite{xue2019video,ehret2019model}, kernel-based methods~\cite{mildenhall2018burst,xu2019learning,marinvc2019multi,xia2019basis} and feature-based alignment~\cite{liu2017robust,tian2018tdan,wang2019edvr,yue2020supervised}.

Pre-alignment methods mostly employ optical flow to estimate the motions and perform warping to compensate for temporal offset. The frame-to-frame method~\cite{ehret2019model} utilizes the TV-$L_1$ algorithm~\cite{zach2007duality} to estimate optical flow within frames. ToFlow~\cite{xue2019video} utilizes the SpyNet~\cite{ranjan2017optical} as the flow estimation module, which is jointly trained with the denoising module. However, the restoration performance of those methods is largely affected by the accuracy of estimated optical flow, while accurate flow is difficult to obtain especially under large motion and severe noise. Kernel-based methods use convolutional neural networks to predict spatially varying kernels, which perform aligning and denoising simultaneously~\cite{mildenhall2018burst}. Compared with the original KPN~\cite{mildenhall2018burst}, Xu \emph{et al.}~\cite{xu2019learning} and Marinc \emph{et al.}~\cite{xu2019learning} proposed to learn the deformable kernels and multiple kernels. Xia \emph{et al.}~\cite{xia2019basis} proposed to predict a set of global basis kernels and the corresponding mixing coefficients to effectively exploit larger denoising kernels. 

Comparing with the above two categories of alignment methods, performing alignment in feature domain is a more promising strategy and has achieved SOTA performance on video super-resolution tasks. Liu \emph{et al.}~\cite{liu2017robust} proposed to use a localization net to estimate spatial transform parameters from deep feature and directly wrap the feature to align shift. TDAN~\cite{tian2018tdan} and EDVR~\cite{wang2019edvr} were proposed to estimate offset of deformable convolution which is utilized to align the shift in features domain. Comparing with~\cite{liu2017robust} which needs the ground-truth information of spatial transform parameters, TDAN and EDVR do not need such parameters while still achieving SOTA performance. RviDeNet~\cite{yue2020supervised} also utilizes DConv to align multi-scale features for burst denoising. In this work, we perform alignment in feature domain and utilize deformable convolution to implicitly compensate for offsets. Moreover, we design an inter-frame module which not only utilizes multi-scale information, but also considers temporal constraint.

\section{Methods}
\label{method}
\subsection{Problem Specification}
Our goal is to recover a clean full-color image, denoted by $x_{ref}^{rgb}$ , from a burst of real-world noisy CFA images, denoted as $\mathcal{Y} = \{ y_t \}_{t=1}^{N}$. The subscript ``ref" represents the index of reference frame. Usually, the noisy counterpart of $x_{ref}^{rgb}$ is the center frame in $\mathcal{Y}$.

The noise in real-world raw images is signal-dependent~\cite{foi2008practical} due to the photon arrival statistics and the imprecision in readout circuitry. The noise introduced by photon sensing, \emph{i.e.}, shot noise, can be modeled as the Poisson distribution, while the noise introduced in readout circuitry, \emph{i.e.}, read noise, can be modeled by the Gaussian distribution. Denote by $x_t$ the desired clean raw image captured at time $t$. The corresponding noisy raw image $y_t$ can be written as~\cite{foi2008practical}:

\begin{equation}
y_t = x_t + n(x_t, \sigma_s, \sigma_r),
\label{noisemodel}
\end{equation}
where $n(x_t, \sigma_s, \sigma_r) \sim \mathcal{N}(0, \sigma_s x_t + \sigma_r^2)$ and $\sigma_s$ and $\sigma_r$ are the scale parameters for shot noise and read noise, respectively. $\mathcal{N}$ represents Gaussian distribution.
%
%The noise map $m_t$ is defined as the the standard deviation.
%

\subsection{Green Channel Prior}
\label{prior}
As discussed in~\cite{tan2014green}, the CMOS sensor has different sensitivity to light of different wavelengths or colors, and in most illumination conditions, green channels are brighter than red and blue channels in Bayer pattern CFA images. Since the real-world noise contains the Poissonian shot noise (see Eq.~\ref{noisemodel}), and the signal-to-noise ratio (SNR) has a square root relationship between signal and noise, the brighter green channel often has a higher SNR than red/blue channels. To validate this, we compute the average SNR of different color channels of 50 real-world noisy raw images randomly chosen from the SIDD~\cite{abdelhamed2018high} benchmark dataset (which contains high-ISO noisy images captured by smartphone cameras), and show the SNR comparison in Fig.~\ref{fig:noise_snr}. One can see that the SNR of green channel is higher than that of red/blue channels for most of the noisy images. In addition, the green channel has twice the sampling rate of red/blue channels. Overall, the green channel preserves better image structure and details than the other two channels. In this paper, we call the prior knowledge that the green channel has higher SNR, higher sampling rate and hence better channel quality than red/blue channels the \emph{green channel prior} (GCP), which is carefully exploited in this paper to design our JDD-B network.

\subsection{Network Structure}
With GCP, we propose a new network, namely GCP-Net, for JDD-B. Without loss of generality, we assume that the Bayer CFA pattern is used. For each raw image $y_t$, we reshape it as four R, G, G, B sub-images of the same size so that $y_t \in \mathbb{R}^{H\times W\times 4}$. We denote the noise map of $y_t$ as $m_t \in \mathbb{R}^{H\times W\times 4}$, whose value at each location is the standard deviation of signal-dependent noise at that position. The input of our GCP-Net is a sequence of noisy raw images $\mathcal{Y} = \{ y_t \}_{t=1}^{N}$ and their corresponding noise maps $\mathcal{M} = \{ m_t \}_{t=1}^{N}$. The output is the clean full-resolution linear RGB image $\hat{x}_{ref}^{rgb} \in \mathbb{R}^{2H\times 2W\times 3}$. 

The overview of GCP-Net is illustrated in Fig.~\ref{figFlow}. GCP-Net consists of two branches, \emph{i.e.}, a GCP branch and a reconstruction branch. In the GCP branch, the green features $f_t^g$ are extracted from the concatenation of noisy green channels, denoted by $y_t^g$, and their noise level maps, denoted by $m_t^g$. This process can be written as:
\begin{equation}
    f_t^g = M_g(y_t^g, m_t^g),
\end{equation}
where $M_g$ consists of several Conv+LReLU blocks. These GCP features are utilized as the guided information for the reconstruction branch. We utilize layer-wise guiding strategy and denote the GCP feature of the $l$-th layer as $f_{t,l}^g$. 

The reconstruction branch utilizes the burst images, the corresponding noise maps and the GCP features to estimate the clean full color image. As illustrated in Fig.~\ref{figFlow}, it consists of three parts: the intra-frame (IntraF) module, the inter-frame (InterF) module and the merge module. The IntraF module is designed to model the deep features of each frame and it utilizes GCP features to guide the feature extraction. The InterF module is to compensate for the shift between frames by using the DConv in feature domain. To reduce the influence of noise in alignment, offset is estimated from the cleaner GCP features. The merge module is designed to aggregate the aligned features and use the GCP features to perform adaptive upsampling for the full-resolution image reconstruction. The details of these modules are presented in the following sections.
\subsection{Intra-frame Module}
\begin{figure}[!t]
\centering
%\hspace{-0.6cm}
\begin{overpic}[width=0.48\textwidth]{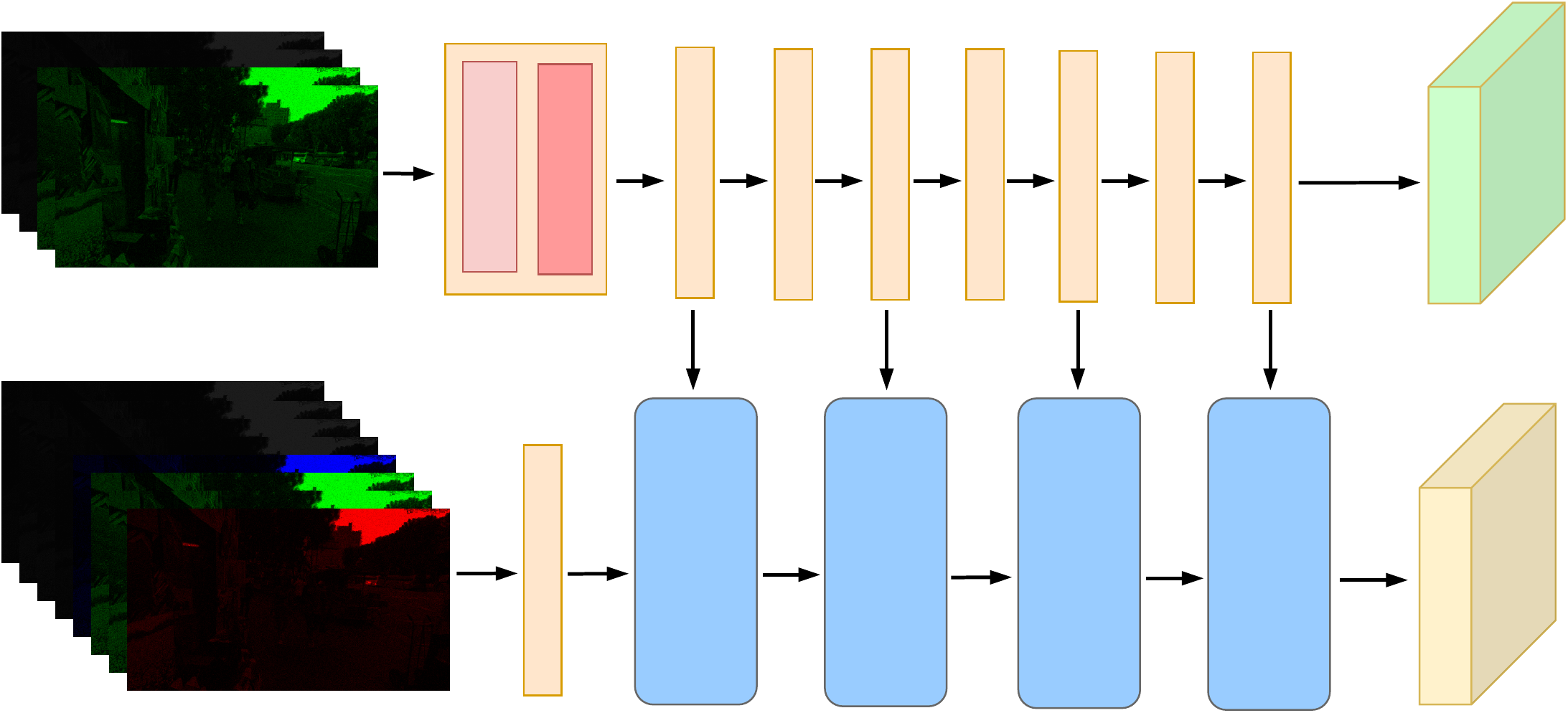}
	\put(10,-1){\color{black}{\footnotesize $y_t, m_t$}}
	\put(10,25){\color{black}{\footnotesize $y_t^g, m_t^g$}}
	\put(30.2,31){\color{black}{\rotatebox{90}{\footnotesize Conv}}}
	\put(35.2,30){\color{black}{\rotatebox{90}{\footnotesize LReLU}}}
	
    \put(43.5,2.4){\color{black}{\footnotesize \rotatebox{90}{\textbf{GCA Block}}}}
	\put(55.5,2.4){\color{black}{\footnotesize \rotatebox{90}{\textbf{GCA Block}}}}
	\put(67.5,2.4){\color{black}{\footnotesize \rotatebox{90}{\textbf{GCA Block}}}}
	\put(80,2.4){\color{black}{\footnotesize \rotatebox{90}{\textbf{GCA Block}}}}
	
	\put(96,0){\color{black}{\footnotesize $f^t$}}
	\put(96,25){\color{black}{\footnotesize $f_t^g$}}
\end{overpic}

\caption{Architecture of the Intra-frame (IntraF) module.}
\label{figIntra}
%\vspace{-0.3cm}
\end{figure}

The architecture of the IntraF module is shown in Fig.~\ref{figIntra}. For the $t$-th frame, the input of IntraF includes the noisy raw image $y_t$, the corresponding noise level map $m_t$ and the GCP feature $f_t^g$. Firstly, one simple convolution layer $M_0$ is used to model the initial features $f_t^0$ as $f_t^0= M_0 (y_t,m_t)$. Then, the initial features $f_t^0$ are passed to the concatenation of four green channel attention (GCA) blocks, where the GCP features are used to guide the feature extraction and a dual attention mechanism is designed to better deal with the channel-dependent and spatial-dependent noise. We adopt a layer-wise guiding strategy for GCA blocks and empirically find that such a strategy is favorable to the restoration results. Without loss of generality, we use the $l$-th GCA block to present the modeling process. The output features are denoted as:
\begin{equation}
    f_{t,l}^{'} = M_{gca}^l (f_{t,l}, f_{t,l}^{g}),
\end{equation}
where $M_{gca}^l$ represents the $l$-th GCA block.

The detailed structure of the GCA block is shown in Fig.~\ref{figGCA}. Inspired by~\cite{he2010guided,park2019semantic}, the GCP information can be exploited by using pixel-wise scaling and bias, and the enhanced feature $f_{t,l}^e$ can be expressed as:
\begin{equation}
    f_{t,l}^e = \gamma(f_{t,l}^{g}) \cdot f_{t,l} + \beta(f_{t,l}^{g}),
\label{gg_equ}
\end{equation}
where $\gamma(f_{t,l}^{g})$ and $\beta(f_{t,l}^{g})$ are two learned modulation parameters of the guided layers. We denote the unit to implement Eq.~\ref{gg_equ} as the green guided (GG) unit. The green-guided features are estimated by two residual blocks, denoted by $M_r$,
\begin{equation}
    f_{t,l}^r = M_r(f_{t,l}^e),
\end{equation}
where $f_{t,l}^r$ is the learned features. As normal Conv layers treat spatial and channel features equally, it is not appropriate to handle the real-world noise which is channel and spatial dependent. To further enhance the representational power of standard Conv+ReLU blocks, channel attention and spatial attention~\cite{hu2018squeeze,anwar2019real,zamir2020learning} are designed to model the cross-channel and spatial relationship of deep features.

%\textbf{Channel attention}: 
\subsubsection{Channel attention (CA)} 
The features of size $H\times W\times C$ are firstly converted into a $1\times 1\times C$ channel descriptor $z_c$ using global average pooling (GAP). To make use of the aggregated information, the channel descriptor is processed by two convolutional layers with kernel size $1\times 1$, followed by a sigmoid activation to obtain the activations $z_c^{'}$. The output of CA is the rescaled feature $f_{t,l}^r$ using $z_c^{'}$.

\subsubsection{Spatial attention (SA)}
The SA block is designed to model the spatial dependencies of deep features by rescaling the features using the estimated spatial attention map $z_s\in \mathbb{R}^{H \times W \times 1}$. Instead of using average pooling and max pooling~\cite{zamir2020learning}, $z_s$ is adaptively obtained by using two convolutional layers, followed by the sigmoid activation.

The output of the $l$-th GCA block is obtained by:
\begin{equation}
    f_{t,l}^{'} = f_{t,l}^r + f_{t,l}^r\cdot z'_c + f_{t,l}^r\cdot z_s.
\end{equation}

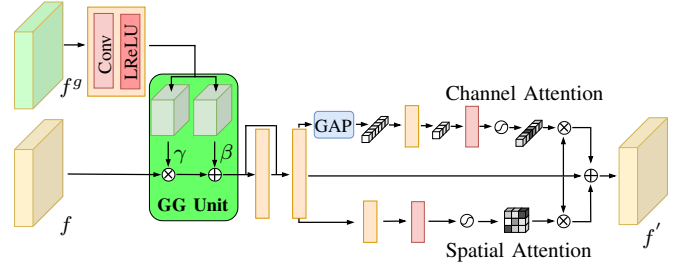
\begin{figure}[!t]
\centering
%\hspace{-0.6cm}
\begin{overpic}[width=0.48\textwidth]{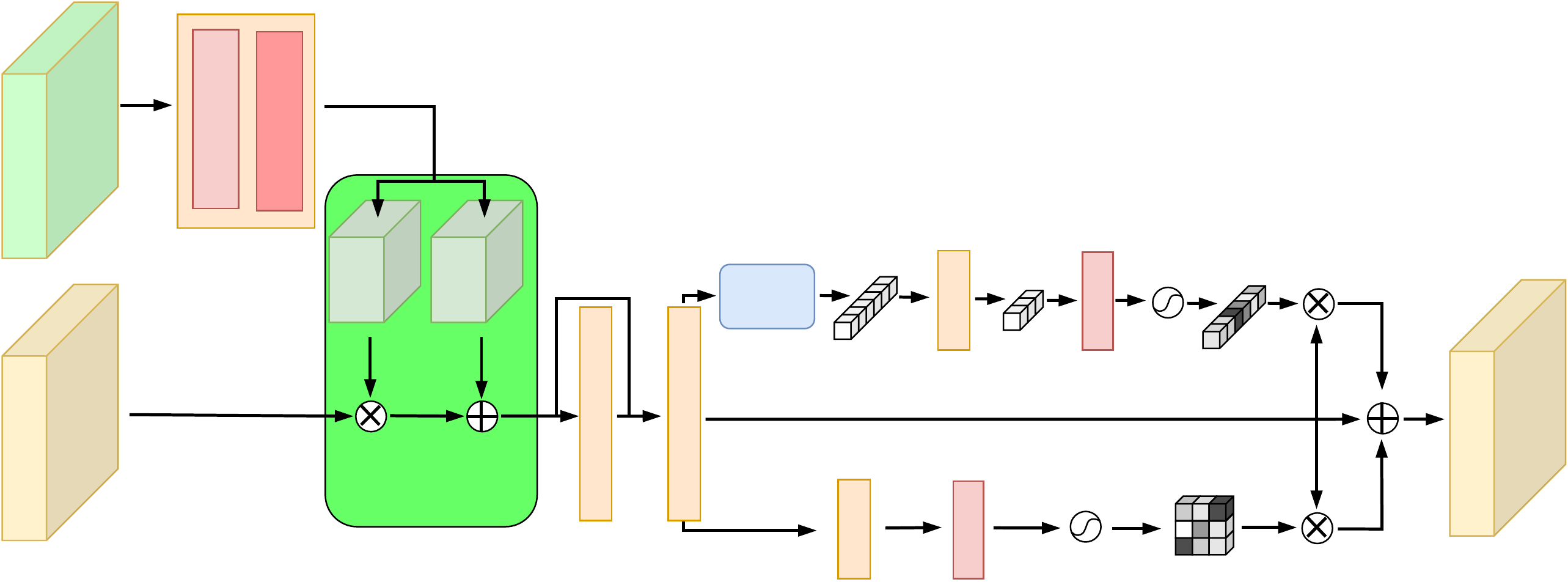}
	\put(7,2){\color{black}{\footnotesize $f$}}
	\put(7,23){\color{black}{\footnotesize $f^g$}}
	\put(13,26){\color{black}{\rotatebox{90}{\scriptsize Conv}}}
	\put(17,25){\color{black}{\rotatebox{90}{\scriptsize LReLU}}}
	
	\put(24.5,13){\color{black}{\footnotesize $\gamma$}}
	\put(31.5,13){\color{black}{\footnotesize $\beta$}}
	\put(22,5){\color{black}{\scriptsize \textbf{GG Unit}}}

	\put(46,17){\color{black}{\scriptsize GAP}}
	\put(66,22){\color{black}{\footnotesize Channel Attention}}
	\put(66,-2){\color{black}{\footnotesize Spatial Attention}}
	\put(96,0){\color{black}{\footnotesize $f^{'}$}}
	
\end{overpic}

\caption{Architecture of the GCA block.}
\label{figGCA}
%\vspace{-0.3cm}
\end{figure}

\subsection{Inter-frame Module}
\begin{figure}[!h]
\centering
%\hspace{-0.6cm}
\begin{overpic}[width=0.35\textwidth]{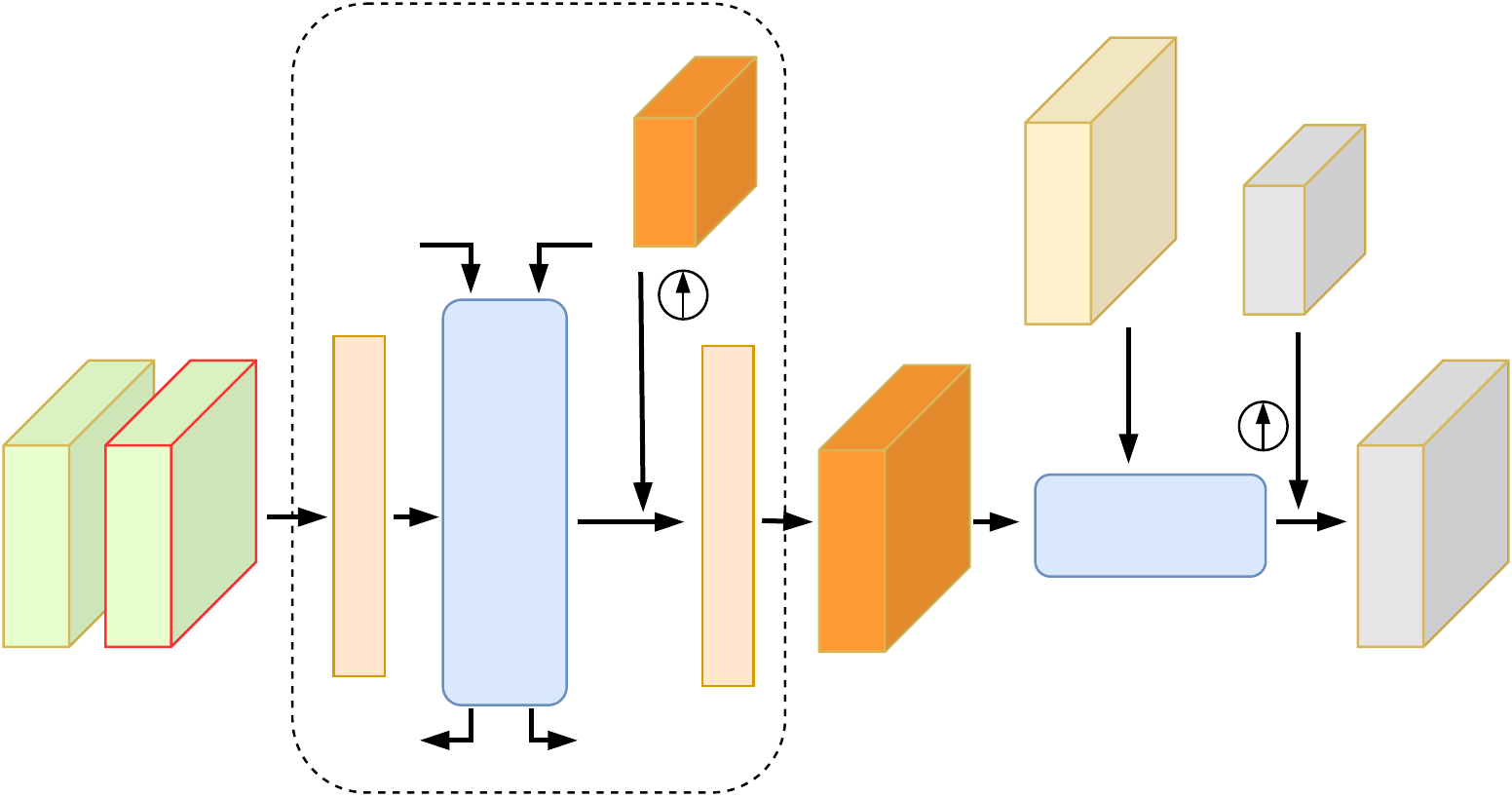}
	\put(0,4.5){\color{black}{\footnotesize $f_t^{g,s}$}}
	\put(8,4.5){\color{black}{\footnotesize $f_{ref}^{g,s}$}}
	\put(70.5,17){\color{black}{\scriptsize \textbf{DConv}}}
	
	\put(25.4,38.5){\color{black}{\scriptsize $h_{t-1}^{s}$}}
	\put(34.2,38.5){\color{black}{\scriptsize $c_{t-1}^{s}$}}
	\put(23.6,1.5){\color{black}{\scriptsize $h_{t}^{s}$}}
	\put(38.2,1.5){\color{black}{\scriptsize $c_{t}^{s}$}}
	\put(32.0,8.5){\color{black}{\footnotesize \rotatebox{90}{\textbf{ConvLSTM}}}}
	
	\put(54,5){\color{black}{\footnotesize $\Delta p_t^{s}$}}
	\put(37,49.5){\color{black}{\tiny $\Delta p_t^{s-1}$}}
	\put(67,49.5){\color{black}{\footnotesize $f_t^{s}$}}
	\put(82,45.5){\color{black}{\footnotesize $f_t^{a,s-1}$}}
	\put(89,4){\color{black}{\footnotesize $f_t^{a,s}$}}
	% \put(22,47){\color{black}{\footnotesize $M_{es}^{s}$}}
\end{overpic}
\caption{Architecture of the Inter-frame (InterF) module for frame $t$ with scale $s$.}
\label{figInter}
%\vspace{-0.3cm}
\end{figure}
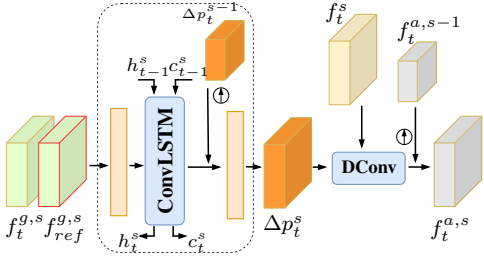

The extracted features $\{ f_t \}_{t=1}^{N}$ by the IntraF module are then aligned to the reference frame feature $f_{ref}$ in the InterF module. The InterF module aims at modeling the temporal dependency between frames, whose architecture is shown in Fig.~\ref{figInter}. We use the deformable convolution to compensate for the offset within frames. To relieve the affect of severe noise and to better model the correlation between neighboring frames, we use the GCP features $f_t^g$ to estimate the offset. Similar to EDVR~\cite{wang2019edvr} and RViDeNet~\cite{yue2020supervised}, pyramidal processing is utilized to handle possible large motions. Moreover, to better exploit the temporal constraint in the offset estimation, we introduce an LSTM regularization in the offset estimation.

For each pyramidal scale $s$ of the $t$-th frame, the inter-frame GCP feature, denoted by $f_{t,i}^s$, is obtained by using
\begin{equation}
f_{t,i}^s = g([f_t^{g, s}, f_{ref}^{g, s}]),
\end{equation}
where $[\cdot,\cdot]$ is the concatenation operator and $g$ is the Conv layer. Then, the temporal regularization is introduced by using ConvLSTM~\cite{xingjian2015convolutional}, which is a popular 2D sequence data modeling method. The ConvLSTM updates the hidden state $h_t^s$ and the cell state $c_t^s$ with:
\begin{equation}
h_t^s, c_t^s = \text{ConvLSTM}(f_{t,i}^s, h_{t-1}^s, c_{t-1}^s).
\end{equation}
The updated inter-frame feature $f_{t,i}^s$ can be written as:
\begin{equation}
f_{t,i}^s = \text{LReLU}(g([f_{t,i}^s, h_t^s])).
\end{equation}
As discussed in~\cite{xiang2020zooming}, the LSTM mechanism has limited ability to deal with complex motions. To handle large motions, multi-scale information is aggregated to estimate more accurately the offset $\Delta p_t^s$:
\begin{equation}
\Delta p^{s}_{t} = g_1(\text{LReLU}(g_2([f_{t,i}^s, (\Delta p^{s-1}_{t})^{\uparrow2}])),
\end{equation}
where $g_1$ and $g_2$ are two convolutional layers, $(\cdot)^{\uparrow 2}$ is the upsampling operator with factor 2.

The aligned features $f_{t}^{s}$ at each position $p_0$ can then be obtained by:
\begin{equation}
\hat{f}_{t}^{s} = \text{DConv}(f_t^s, \Delta p_t^s) = \sum_{k=1}^K w_k \cdot f_{t}(p_0 + p_k + \Delta p_{t,k}^s) \cdot \Delta m_k,
\label{deforma}
\end{equation}
in which $K$ is the sampling location of deformable convolution kernel, $\Delta m_t^s$ is the modulation scalar. 
Following \cite{dai2017deformable}, since $\Delta p_t^s$ is fractional, bilinear interpolation is applied. The final aligned feature $f_{t}^{a,s}$ at scale $s$ is obtained by:
\begin{equation}
f_{t}^{a,s} = M_g(\hat{f}_{t}^{s}, (\hat{f}_{t}^{s-1})^{\uparrow2}),
\label{pyramidal3}
\end{equation}
where $M_g$ refers to general Conv+LReLU layers.

\subsection{Merge Module}
The merge module is designed to merge the aligned features $f_t^a$ and output the estimated clean RGB image $\hat{x}_{ref}^{rgb}$. The aligned features are firstly concatenated and adaptively merged as follows:
\begin{equation}
f^m = M_m(\{f_t^a\}_{t=1}^N),
\label{mergefun}
\end{equation}
where $M_m$ is the merge function by using one simple convolution layer. Then we upsample the features $f^m\in \mathcal{R}^{H\times W\times C}$ to full-resolution features $f^m_f\in \mathcal{R}^{2H\times 2W\times C}$ using GCP adaptive upsampling. Similar to the green guided operator in GCA block, the GCP adaptive upsampling can be expressed as:
\begin{equation}
    f^m_f = \gamma((f_g)^{\uparrow 2})\cdot(f^m)^{\uparrow 2} + \beta((f_g)^{\uparrow 2}),
\end{equation}
where $f_g$ are the green guided features of size $H \times W \times C$, $(\cdot)^{\uparrow s}$ represents the upsampling operator by a factor $s$, $\gamma(f_g)$ and $\beta(g_g)$ are the two learned modulation parameters. Transpose convolution is used for the upsampling interpolation.

The final estimation $\hat{x}_{ref}^{rgb}$ can be written as:
\begin{equation}
    \hat{x}_{ref}^{rgb} = M_U(f^m_f),
\end{equation}
where $M_U$ is designed to estimate the final clean RGB image. In this paper, we utilize a three scale U-Net~\cite{ronneberger2015u} architecture for $M_U$ to exploit the multi-scale information as well as enlarging the receptive field. All the Conv kernels are of size $3\times3$, followed by the nonlinear function LReLU. The upsampling and downsampling operators in $M_U$ are strided convolutions and transpose convolutions.
\subsection{Loss Function}
For the estimated clean $\hat{x}_{ref}^{rgb}$, we define the reconstruction loss in the linear color space as follows:
\begin{equation}
\mathcal{L}_{linear} = \sqrt{\Vert \hat{x}_{ref}^{rgb} - x_{ref}^{rgb} \Vert^2 + \epsilon^2},
\label{llinear}
\end{equation}
where $\sqrt{(\hat{x} - x)^2 + \epsilon^2}$ is the Charbonnier penalty function~\cite{lai2017deep}, $\epsilon$ is set to $1\times 10^{-3}$.
As discussed in KPN~\cite{mildenhall2018burst}, computing loss in sRGB color space can produce a perceptually more relevant estimation. Therefore, we also introduce a loss in the sRGB color space:
\begin{equation}
\mathcal{L}_{srgb} = \sqrt{\Vert \Gamma(\hat{x}_{ref}^{rgb}) - \Gamma(x_{ref}^{rgb}) \Vert^2 + \epsilon^2},
\label{lsrgb}
\end{equation}
where $\Gamma(\cdot)$ is the operator which transforms linear RGB color space to sRGB space. In this paper, $\Gamma(\cdot)$ contains white balance, color correction and gamma compression as in \cite{brooks2019unprocessing}. To sum up, the overall loss to optimize our model is: 
\begin{equation}
\mathcal{L} = \mathcal{L}_{linear} + \lambda\mathcal{L}_{srgb},
\label{lossfunction}
\end{equation}
where $\lambda$ is the trade-off parameter and we simply set to 1 in our experiments.

\section{Experiment}
\label{exp}
\subsection{Implement Details}
\label{implementdetails}
Obtaining ground-truth images for training is difficult for real-world image restoration tasks. In some single-image based restoration works~\cite{plotz2017benchmarking,abdelhamed2018high,qian2019trinity}, real-world degraded data are collected and the corresponding ground-truth images are physically and/or mathematically estimated for pair-wise training. However, for the burst-images based JDD-B task, the misalignment problem and the coherence between denoising and demosaicking make the ground-truth estimation much more difficult. Therefore, we synthesize training data by using an open high quality video dataset, \emph{i.e.}, Vimeo-90K~\cite{xue2019video}.

Since camera sensor outputs are in the linear color space, we first convert the sRGB images $x$ into linear RGB space $x^{rgb}$ by using the unprocessing operation in \cite{brooks2019unprocessing}, which can be written as $x^{rgb} = \Gamma^{-1}(x)$. The unprocessing operation $\Gamma^{-1}(\cdot)$ includes inverse gamma compression, inverse color correction, inverse tone mapping and inverse white balance. The converted linear RGB frame $x^{rgb}$ is taken as the ground-truth image. By using Eq.~\ref{noisemodel}, the noisy raw image $y_t$ can be synthesized as:
\begin{equation}
y_t = M(x_t^{rgb}) + n(M(x_t^{rgb}), \sigma_s, \sigma_r),
\label{noisemodel2}
\end{equation}
where $M(\cdot)$ is the mosaic matrix which downsamples a linear RGB image to a Bayer CFA image. Without loss of generality, the RGGB mosaic pattern is used as $M(\cdot)$ to generate the data.

\begin{table*}[!tbp]
\setlength{\abovecaptionskip}{0.2cm}
\setlength{\belowcaptionskip}{-0.cm}
\footnotesize
%\scriptsize
\centering
\smallskip
\caption{The JDD results of different variants of GCP-Net on the \textbf{REDS4} dataset.}
\label{dis_gg}
%\begin{tabularx} {\linewidth} {P{1.5cm} P{1.3cm} P{1.3cm} %P{1.3cm} P{1.3cm} P{1.3cm} P{1.3cm} P{2.8cm} P{2.0cm}}
%\toprule
%\# of GG Units  & 0 & 1 & 2 & 3 & 4 & 5 & 4 (w/o GCP upsampling) %& using RB to guide \\
%\midrule
%\emph{Clip000} &32.02/0.8976 &32.15/0.9003 &32.14/0.8997 %&32.15/0.8999 &32.29/0.9035  &\textbf{32.31/0.9036} %&\centering32.10/0.8999 &30.98/0.8645\\ 
%\emph{Clip011} &33.14/0.8769 &33.26/0.8778 &33.36/0.8799 %&33.26/0.8789 &\textbf{33.28/0.8783} &33.27/0.8782 %&\centering33.22/0.8768 &32.79/0.8697 \\
%\emph{Clip015} &35.64/0.9159 &35.79/0.9176 &35.83/0.9184 %&35.80/0.9182 &\textbf{35.79/0.9177} &\textbf{35.79/0.9177} %&\centering35.74/0.9170 &35.18/0.9085 \\
%\emph{Clip020} &32.48/0.8958 &32.59/0.8974 &32.64/0.8985 %&32.61/0.8981 &\textbf{32.63/0.8979} &32.62/0.8978 %&\centering32.60/0.8979 &32.11/0.8890 \\
%Average &33.32/0.8966 &33.45/0.8983 &33.49/0.8991 &33.46/0.8988 %&\textbf{33.50/0.8994} &33.50/0.8993 &\centering33.41/0.8979 %&32.77/0.8829 \\
%\toprule
\begin{tabularx} {\linewidth} {P{1.8cm} P{1.1cm} P{1.1cm} P{1.1cm} P{1.1cm} P{1.1cm} P{1.1cm} P{3.4cm} P{2.2cm}}
\toprule
\# of GG Units  & 0 & 1 & 2 & 3 & 4 & 5 & 4 (w/o GCP upsampling) & using RB to guide \\
\midrule
\emph{Clip000} &32.02 &32.15 &32.14 &32.15 &32.29  &\textbf{32.31} &32.10 &30.98\\ 
\emph{Clip011} &33.14 &33.26 &33.36 &33.26 &\textbf{33.28} &33.27 &33.22 &32.79 \\
\emph{Clip015} &35.64 &35.79 &35.83 &35.80 &\textbf{35.79} &\textbf{35.79} &35.74 &35.18 \\
\emph{Clip020} &32.48 &32.59 &32.64 &32.61 &\textbf{32.63} &32.62 &32.60 &32.11 \\
Average &33.32 &33.45 &33.49 &33.46 &\textbf{33.50} &33.50 &33.41 &32.77 \\
\toprule
\end{tabularx} 
\end{table*}

In our experiments, a number of $N=5$ neighboring frames are used as the input and the central frame is chosen as the reference frame. Following the setting in \cite{mildenhall2018burst}, the noise level parameters $\sigma_s$ and $\sigma_r$ in Eq.~\ref{noisemodel} are uniformly sampled from the ranges of $[10^{-4}, 10^{-2}]$ and $[10^{-3}, 10^{-1.5}]$, respectively. We adopt the method in \cite{he2015delving} to initialize the GCP-Net and use the ADAM~\cite{kingma2014adam} algorithm with $\beta_1$ = 0.9 and $\beta_2$ = 0.99 to update the network. The size of mini-batch is 2 and the size of each noisy raw patch is $64\times64$ with 4 color channels (RGGB). The reconstructed RGB patch is of size $128\times128$ with three color channels (RGB). The learning rate is initialized as $4\times 10^{-4}$ and it is decreased using the cosine function~\cite{loshchilov2016sgdr}. It takes about two days to train our model under the PyTorch framework using two Nvidia GeForce RTX 2080 Ti GPU.

\subsection{Ablation Study}
\label{discussion}
In this section, we perform ablation studies to discuss the effect of major components in GCP-Net and the setting of some parameters. The Vid4~\cite{liu2013bayesian} and REDS4~\cite{wang2019edvr} datasets are used in the experiments.

\subsubsection{The effectiveness of GCP}
\label{GCAdis}
\begin{figure}[!h]
\setlength{\abovecaptionskip}{0.2cm}
\centering
\begin{minipage}[b]{0.48\textwidth}
\centering
    \begin{minipage}[b]{0.3\textwidth}
    \centering
    \subfigure[hang,scriptsize,tight][Noisy Image]{
        \includegraphics[width=1\textwidth]{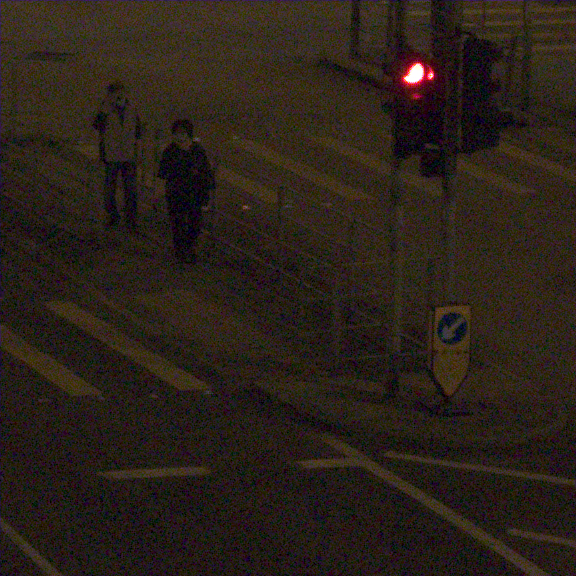}
        }
    % {\footnotesize  (b)}
    \end{minipage}
    \begin{minipage}[b]{0.3\textwidth}
    \centering
    \subfigure[hang,scriptsize,tight][Restored Image]{
        \includegraphics[width=1\textwidth]{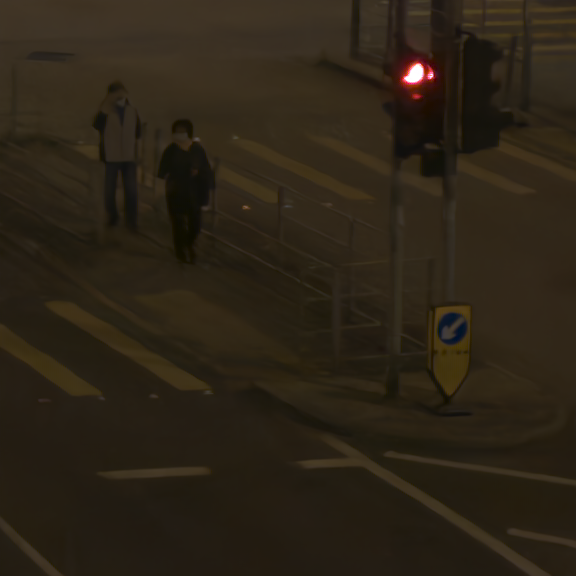}
        }
    % {\footnotesize  (b)}
    \end{minipage}
    \begin{minipage}[b]{0.3\textwidth}
    \centering
    \subfigure[hang,scriptsize,tight][$\gamma$]{
        \includegraphics[width=1\textwidth]{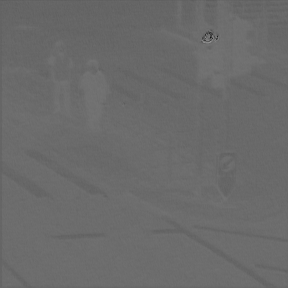}
    }
    % {\footnotesize  w/o GG}
    \end{minipage} 
    %\vspace{0.10cm}
\end{minipage} \hspace{-1ex}
\begin{minipage}[b]{0.48\textwidth}
    \centering
    \begin{minipage}[b]{0.3\textwidth}
    \centering
    \subfigure[hang,scriptsize,tight][$\beta$]{
        \includegraphics[width=1\textwidth]{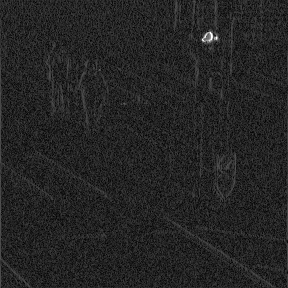}
    }
    \end{minipage} 
    \begin{minipage}[b]{0.3\textwidth}
    \centering
    \subfigure[hang,scriptsize,tight][$f$]{
        \includegraphics[width=1\textwidth]{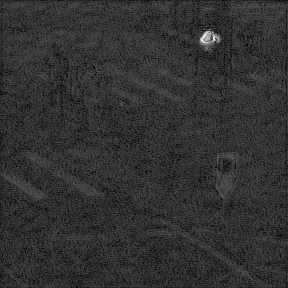}
    }
    % {\footnotesize  w/o GG}
    \end{minipage} 
    \begin{minipage}[b]{0.3\textwidth}
    \centering
    \subfigure[hang,scriptsize,tight][$f'$]{
        \includegraphics[width=1\textwidth]{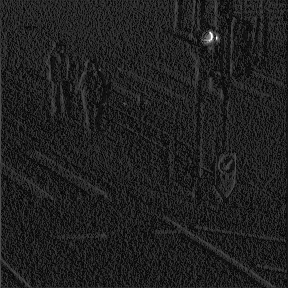}
    }
    % {\footnotesize  w/o GG}
    \end{minipage} 
    
\end{minipage}%\hspace{-2ex}

\caption{JDD results on a real-world noisy image captured by iPhoneX. (a) is a noisy patch and (b) is the restored patch by our GCP-Net. (c) and (d) visualize the $\lambda$ and $\beta$ values (refer to Eq.~\ref{gg_equ}) of the 47-th feature channel. (e) and (f) visualize the extracted deep features without and with the GG unit.}%\vspace{-0.4cm}
\label{figgg}
\end{figure}

In GCP-Net, GCP features are used to guide the deep feature extraction and the upsampling process. By removing and adding the GG unit (see Fig.~\ref{figGCA}) in the GCA block, we can analyze the influence of GCP on deep feature extraction. Fig.~\ref{figgg} shows a patch of a noisy image captured in night time by a smartphone camera, the extracted deep features without and with the GG unit, and the JDD results. As expected, using the GCP features to guide the feature extraction is favorable to suppress the noise and preserve more detailed textures, as shown in Figs.~\ref{figgg} (e) and (f).

To quantitatively verify the contribution of GCP, we implement five variants of GCP-Net with different number of GG units and GCA blocks. The quantitative results on the REDS4 dataset are shown in Table~\ref{dis_gg}. We can see that using one GG unit, we can obtain 0.13dB gain over the result without using the GCP guidance. The PSNR value can be further improved by increasing the number of GG units from one to four, and the performance gets saturated when the number of GG units is five. Therefore, we use four GG units and GCA blocks in our GCP-Net. We also train a GCP-Net without using the adaptive upsampling in GCA and the result is shown in Table~\ref{dis_gg}. One can see that the adaptive upsampling can obtain about 0.1dB gain for the JDD task.

We also train a network by using the red and blue channels to guide the feature extraction. The results are shown in the last column of Table~\ref{dis_gg}. We see that using the red and blue channels to guide feature extraction cannot enhance the performance. Instead, it leads to serious performance degradation (about 0.6dB) compared with the network without using GCP. This is not surprising since the red and blue channels have lower SNR and contain less textures (please see Figs.~\ref{figcover}(d)-(g)) so that the network fails to extract more guiding information to enhance the deep features.

\begin{figure*}[!t]
\centering
\subfigure{
\begin{minipage}[t]{0.24\textwidth}
\centering
\includegraphics[width=1\textwidth]{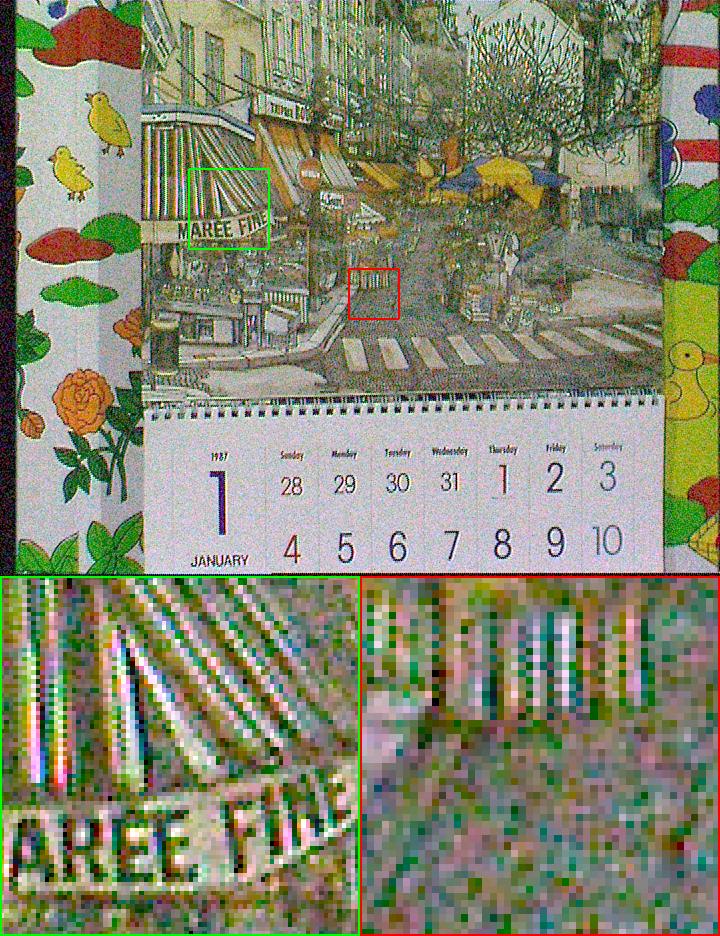}
{\footnotesize  (a) Noisy image}
\end{minipage}\hspace{0.05cm}

\begin{minipage}[t]{0.24\textwidth}
\centering
\includegraphics[width=1\textwidth]{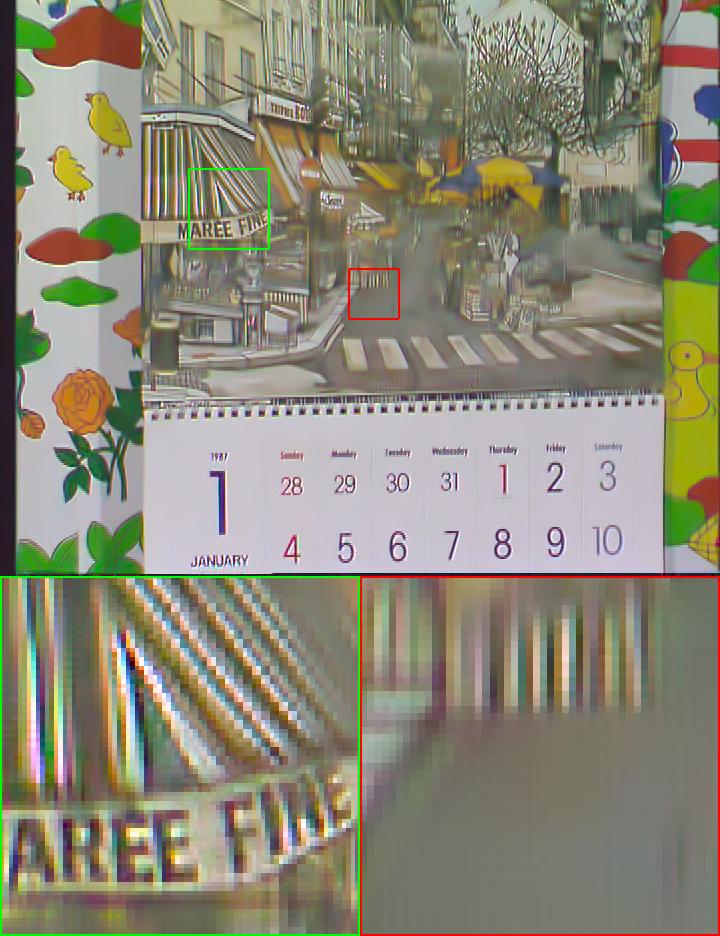}
{\footnotesize  (b) VBM3D+DMN}
\end{minipage}\hspace{0.05cm}

\begin{minipage}[t]{0.24\textwidth}
\centering
\includegraphics[width=1\textwidth]{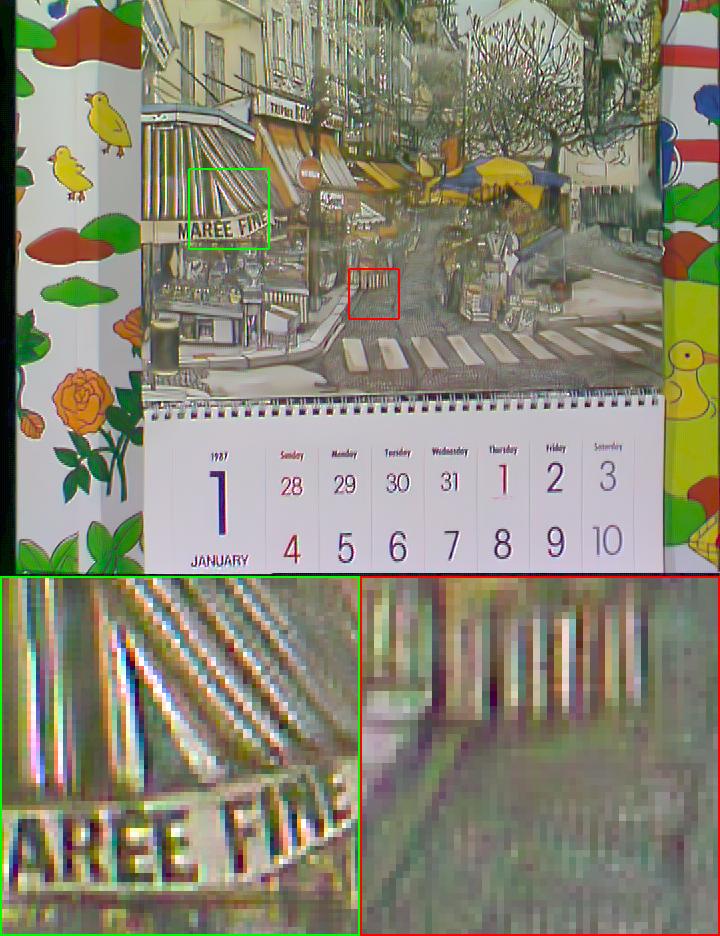}
{\footnotesize  (c) KPN+DMN}
\end{minipage}\hspace{0.05cm}

\begin{minipage}[t]{0.24\textwidth}
\centering
\includegraphics[width=1\textwidth]{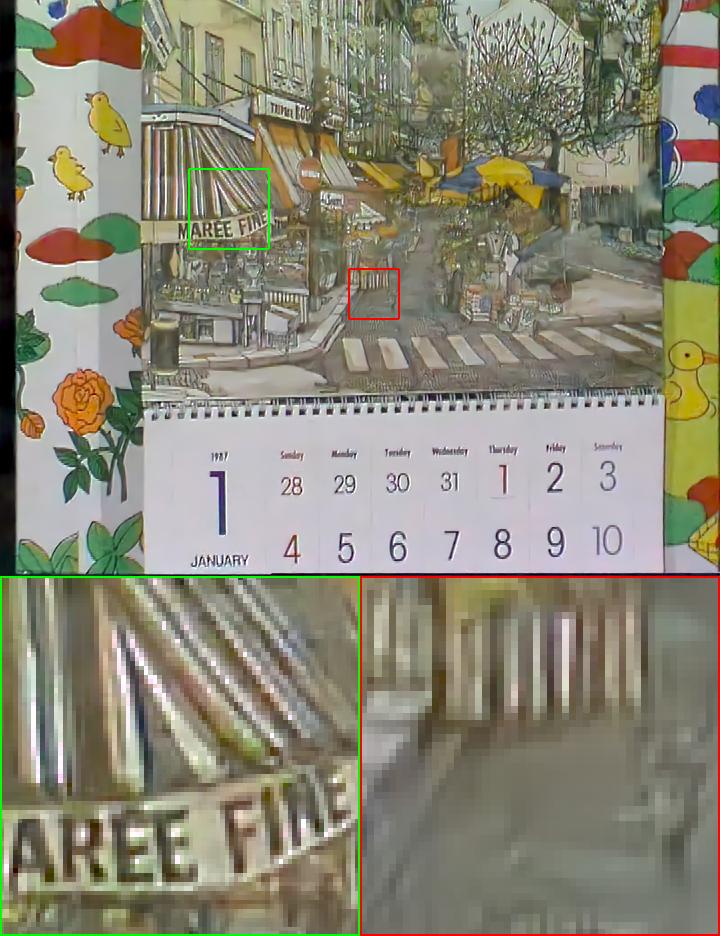}
{\footnotesize  (d) EDVR*}
\end{minipage}
}

%\vspace{-3mm}

\subfigure{
\begin{minipage}[t]{0.24\textwidth}
\centering
\includegraphics[width=1\textwidth]{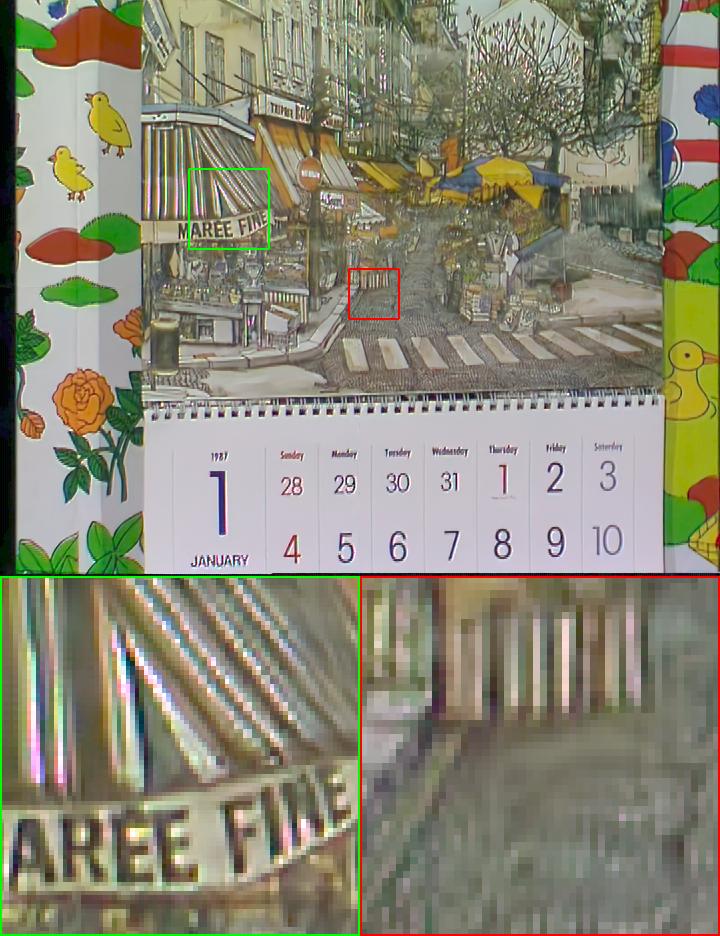}
{\footnotesize  (e) RViDeNet+DMN}
\end{minipage}\hspace{0.05cm}

\begin{minipage}[t]{0.24\textwidth}
\centering
\includegraphics[width=1\textwidth]{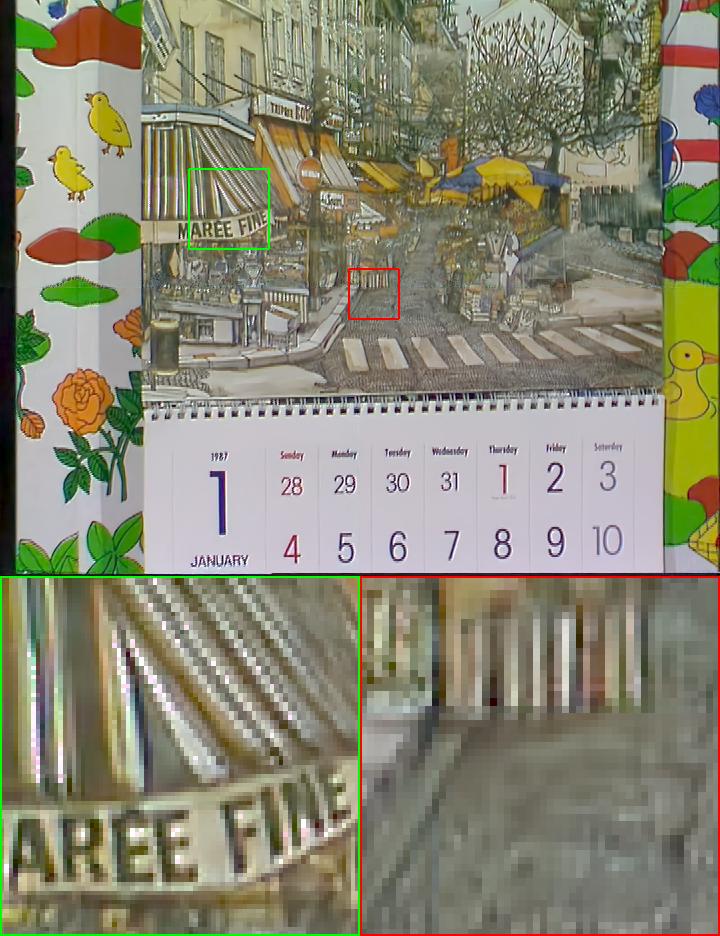}
{\footnotesize  (f) RViDeNet*}
\end{minipage}\hspace{0.05cm}

\begin{minipage}[t]{0.24\textwidth}
\centering
\includegraphics[width=1\textwidth]{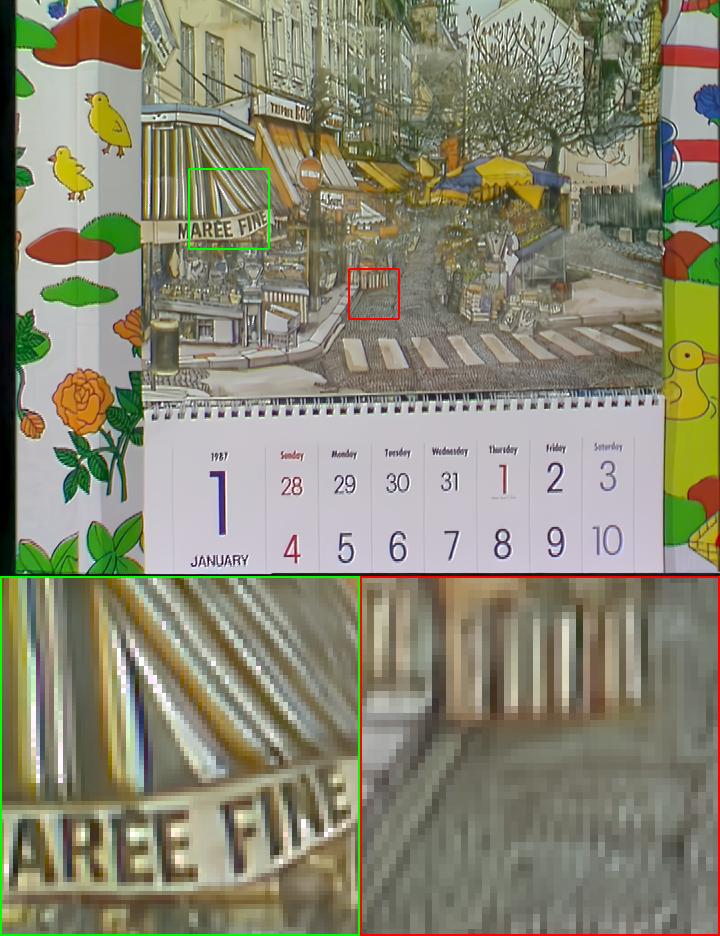}
{\footnotesize  (g) Ours}
\end{minipage}\hspace{0.05cm}

\begin{minipage}[t]{0.24\textwidth}
\centering
\includegraphics[width=1\textwidth]{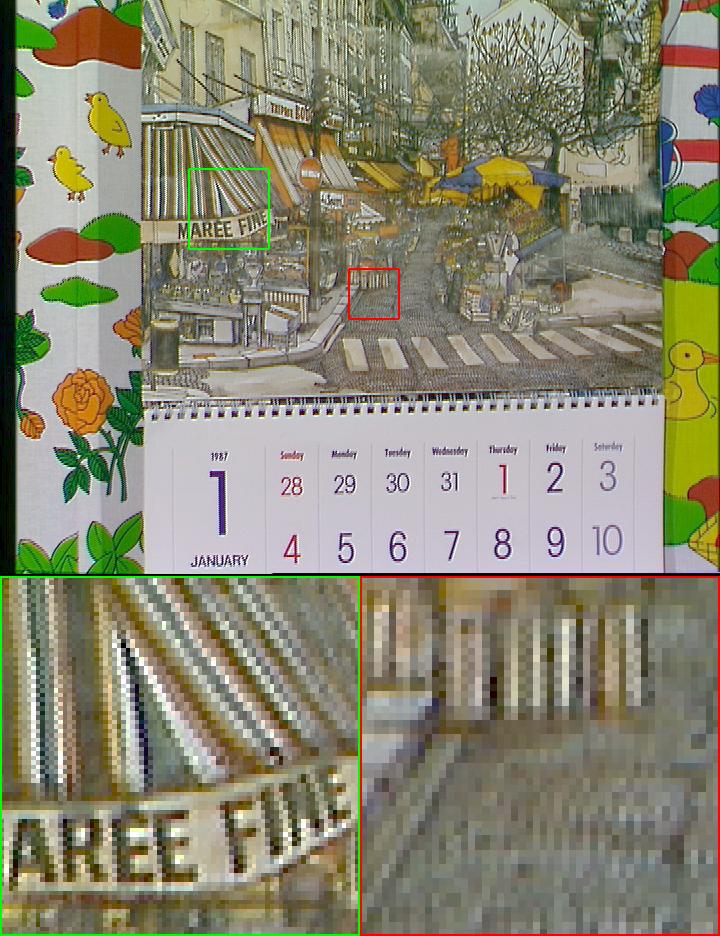}
{\footnotesize  (h) GT}
\end{minipage}
}

\caption{JDD results on the \emph{Calendar} clip in Vid4 dataset by different methods.}%\vspace{-0.4cm}
\label{figVid}
\end{figure*}

\subsubsection{One-stage structure vs. two-stage structure for JDD}
\label{onestage}
There are two types of network structures for JDD: one-stage and two-stage. One-stage algorithms~\cite{henz2018deep,liu2020joint} learn to directly estimate the clean demosaicked image, while two-stage algorithms~\cite{gharbi2016deep,tan2017color,qian2019trinity,ehret2019joint,jin2020review} sequentially learn the denoisng task and the demosaicking task. In this part, we evaluate which structure is more effective to reconstruct the full color images from a burst of noisy mosaic images with similar trainable parameters.

We train two variants of GCP-Net as the two-stage networks for evaluation, namely, GCP-Net-DE+DM and GCP-Net-DM+DE. GCP-Net-DE+DM first performs burst denoising to obtain the clean mosaic image $x_{ref}$ and then applies demosaicking. The intermediate denoising loss is applied on the estimated $\hat{x}_{ref}$, which is obtained by adding one conv layer after the merge function $M_m$ in Eq.~\ref{mergefun}. The denoised image is then taken as the input for the remaining layers to perform demosaicking. For GCP-Net-DM+DE, it firstly performs demosaicking on every noisy raw image and then performs burst denoising on the demosaicked images. The demosaicking output of the $t$-th frame is a noisy demosaicked image $\hat{y}^{rgb}_t$, which is estimated from the output of the last GCA block in the IntraF module. Since there's no ground-truth for $\hat{y}^{rgb}_t$, we pretrain the first stage of GCP-Net-DM+DE for single image demosaicking and then fine-tune the whole network for JDD task. 

Table~\ref{dis_joint} lists the average PSNR results of the variants of GCP-Net on the Vid4 and REDS4 datasets. We can see that our one-stage GCP-Net consistently outperforms its two-stage variants. This is mainly because denoising and demosaicking are two highly relevant tasks and the two-stage network is not effective to exploit the correlation information of these two tasks. Similar to those multi-task learning works~\cite{liu2017wide}, learning simultaneously the relevant tasks could result in better performance. Thus, we choose to use one-stage structure in GCP-Net.

\begin{table}[t]
\setlength{\abovecaptionskip}{0.2cm}
\footnotesize
%\scriptsize
\centering
%\smallskip
\caption{Comparison of the one-stage and two-stage variants of GCP-Net on \textbf{Vid4} and \textbf{REDS4} datasets.}
\label{dis_joint}
%\begin{tabularx}{\linewidth} {c c c c c}
%\begin{tabularx} {\linewidth} {P{0.7cm} P{1.5cm} P{1.5cm} %P{1.5cm} P{1.5cm}}
%\toprule
%Testset &Noise Level &DE + DM & DM + DE & JDD (Ours) \\
%\midrule
%\multirow{2}{*}{Vid4} &High  &31.89/0.9149 &31.67/0.9192 %&\textbf{32.30/0.9205} \\
%&Low  &33.46/0.9394 &33.19/0.9357 &\textbf{33.91/0.9429} \\
%\midrule
%\multirow{2}{*}{REDS4} &High  & 33.00/0.8890 &32.80/0.8853 %&\textbf{33.50/0.8994} \\
%&Low  & 35.05/0.9252 &34.79/0.9209 &\textbf{35.57/0.9318} \\
%\toprule
\begin{tabularx} {\linewidth} {P{0.9cm} P{1.5cm} P{1.5cm} P{1.5cm} P{1.5cm}}
\toprule
Testset &Noise Level &DE + DM & DM + DE & JDD (Ours) \\
\midrule
\multirow{2}{*}{Vid4} &High  &31.89 &31.67 &\textbf{32.30} \\
&Low  &33.46 &33.19 &\textbf{33.91} \\
\midrule
\multirow{2}{*}{REDS4} &High  & 33.00 &32.80 &\textbf{33.50} \\
&Low  & 35.05 &34.79 &\textbf{35.57} \\
\toprule

\end{tabularx} 
\end{table}

\subsubsection{Role of inter-frame module}
To demonstrate the contribution of the proposed inter-frame module (see Fig.~\ref{figInter}), we implement four variants of GCP-Net, \emph{i.e.}, GCP-Net-w/o-GCP, GCP-Net-w/o-inter, GCP-Net-w/o-MS and GCP-Net-w/o-LSTM. Specifically, GCP-Net-w/o-GCP represents the network without using GCP in the inter-frame module. That is, the offset is directly estimated from the original features, instead of GCP features. In GCP-Net-w/o-inter, we remove the inter-frame module and take the concatenation of the output features of IntraF module as the input of the merge module. GCP-Net-w/o-MS is implemented by removing the multi-scale offset estimation in the interF module, and GCP-Net-w/o-LSTM represents the network without temporal regularization in the offset estimation.

Table.~\ref{dis_inter} reports the PSNR results of GCP-Net and its four variants with different interF modules. As expected, full GCP-Net achieves the best performance, showing that compensating for the shift between frames is crucial for burst image restoration. Estimating offset from better quality GCP features can obtain 0.05dB improvement on JDD-B. Compared with GCP-Net-w/o-MS, utilizing multi-scale information benefits to handle large and complex motions, which results in 0.5$\sim$0.6dB improvement on the REDS4 dataset. By introducing temporal regularization in the offset estimation part, our full model further improves the JDD-B results by 0.06dB.

\begin{table}[t]
\setlength{\abovecaptionskip}{0.2cm}
\footnotesize
%\scriptsize
\centering
%\smallskip
\caption{Quantitative comparison of GCP-Nets with different various of InterF on \textbf{REDS4} dataset.}
\label{dis_inter}
%\begin{tabularx}{\linewidth} {c c c c c}
%\begin{tabularx} {\linewidth} {P{0.3cm} P{1.2cm} P{1.2cm} %P{1.2cm} P{1.2cm} P{1.2cm}}
%\toprule
%Noise  & w/o GCP & w/o Inter & w/o MS & w/o LSTM & full \\
%\midrule
% High &33.45/0.8982 &32.69/0.8836 &32.96/0.8894 &33.44/0.8983 & %\textbf{33.50/0.8994}  \\
% Low &35.51/0.9310 &34.80/0.9213 &35.01/0.9246 &35.52/0.9316 & %\textbf{35.57/0.9318} \\
%\toprule

\begin{tabularx} {\linewidth} {P{0.7cm} P{1.1cm} P{1.1cm} P{1.1cm} P{1.3cm} P{1.1cm}}
\toprule
Noise  & w/o GCP & w/o Inter & w/o MS & w/o LSTM & full \\
\midrule
 High &33.45 &32.69 &32.96 &33.44 & \textbf{33.50}  \\
 Low &35.51 &34.80 &35.01 &35.52 & \textbf{35.57} \\
\toprule

\end{tabularx} 
\end{table}

\begin{figure*}[!t]
\centering
\subfigure{
\begin{minipage}[t]{0.24\textwidth}
\centering
\includegraphics[width=1\textwidth]{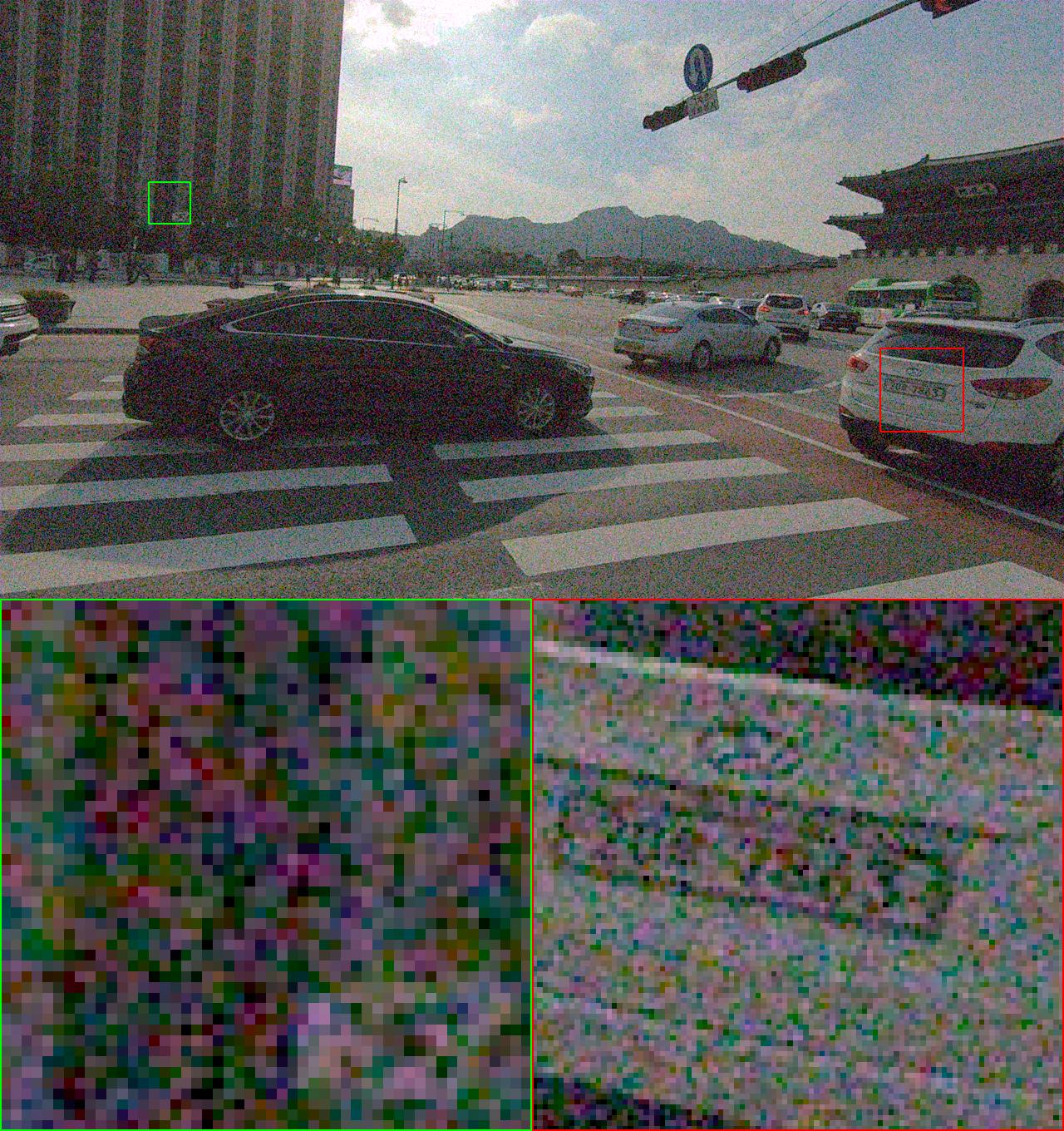}
{\footnotesize  (a) Noisy image}
\end{minipage}\hspace{0.05cm}

\begin{minipage}[t]{0.24\textwidth}
\centering
\includegraphics[width=1\textwidth]{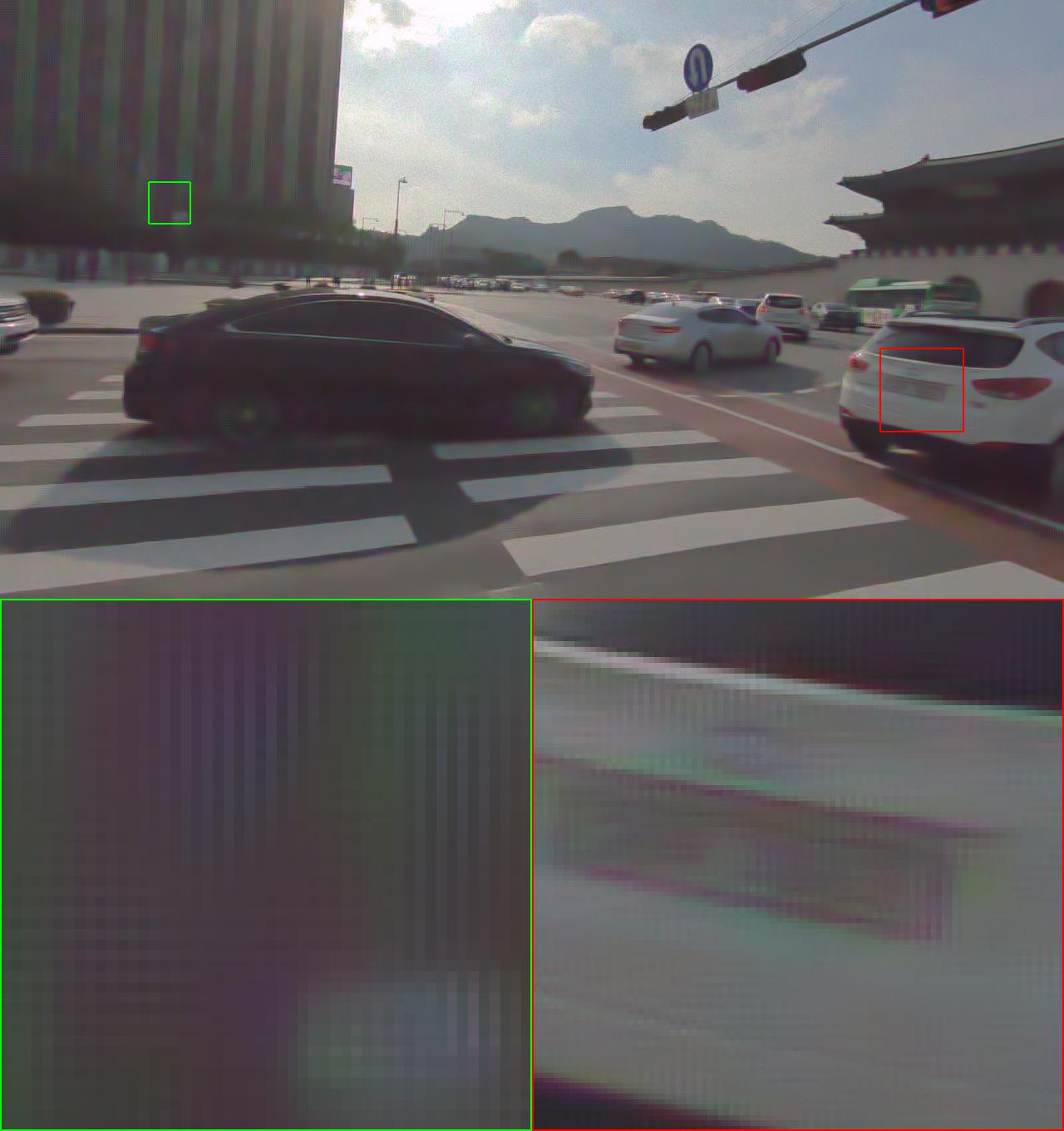}
{\footnotesize  (b) VBM3D+DMN}
\end{minipage}\hspace{0.05cm}

\begin{minipage}[t]{0.24\textwidth}
\centering
\includegraphics[width=1\textwidth]{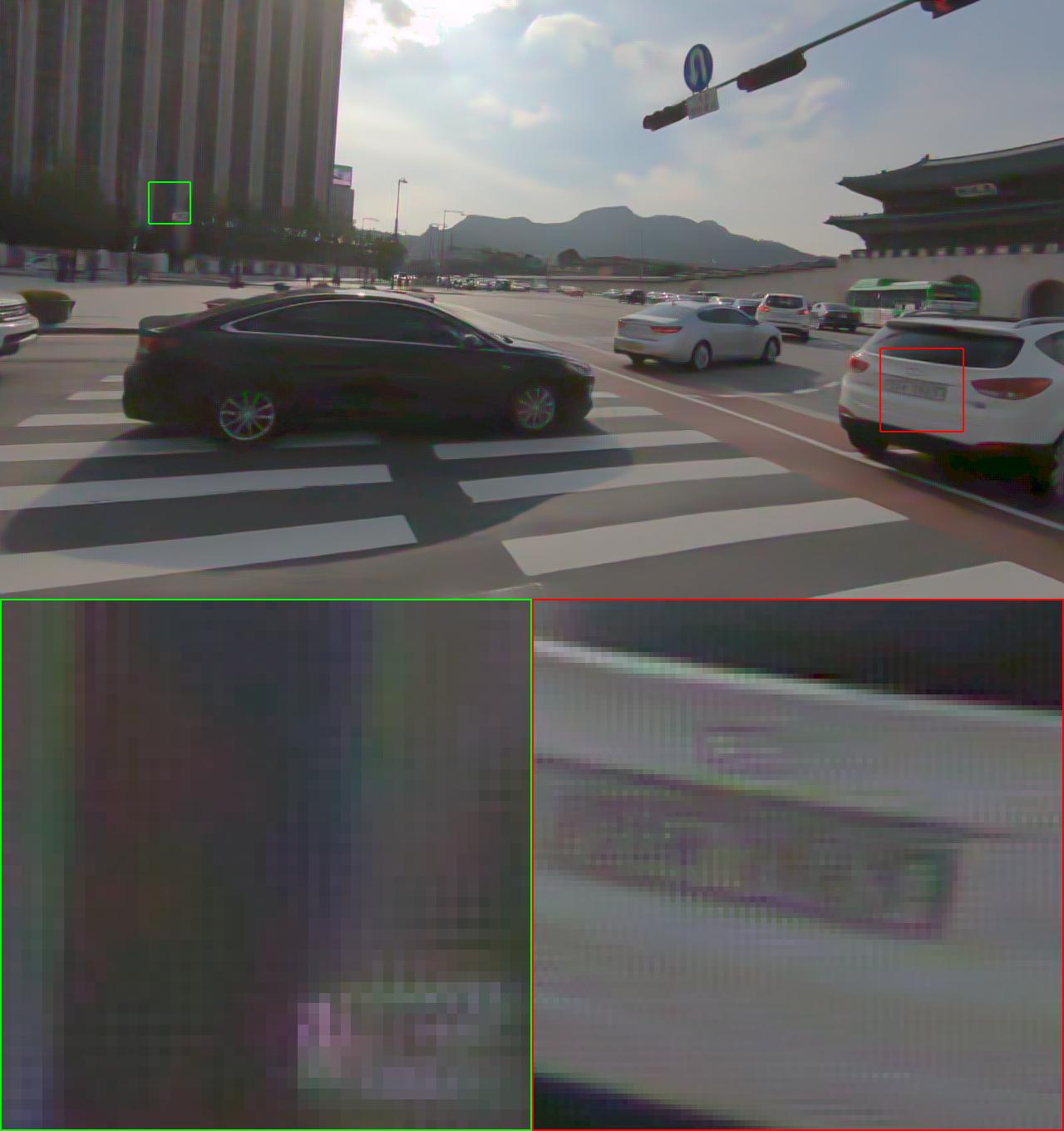}
{\footnotesize  (c) KPN+DMN}
\end{minipage}\hspace{0.05cm}

\begin{minipage}[t]{0.24\textwidth}
\centering
\includegraphics[width=1\textwidth]{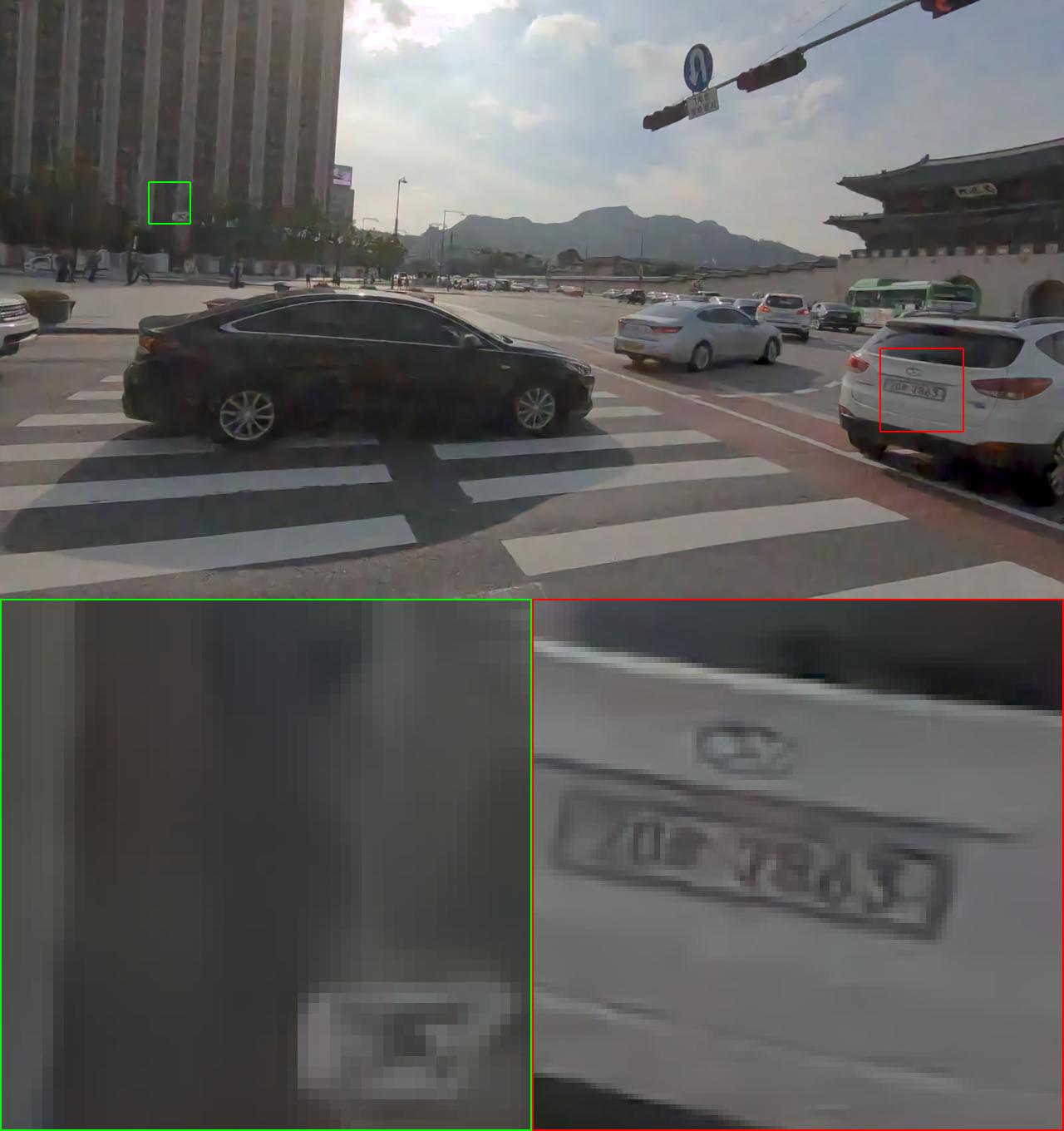}
{\footnotesize  (d) EDVR*}
\end{minipage}
}

%\vspace{-3mm}

\subfigure{
\begin{minipage}[t]{0.24\textwidth}
\centering
\includegraphics[width=1\textwidth]{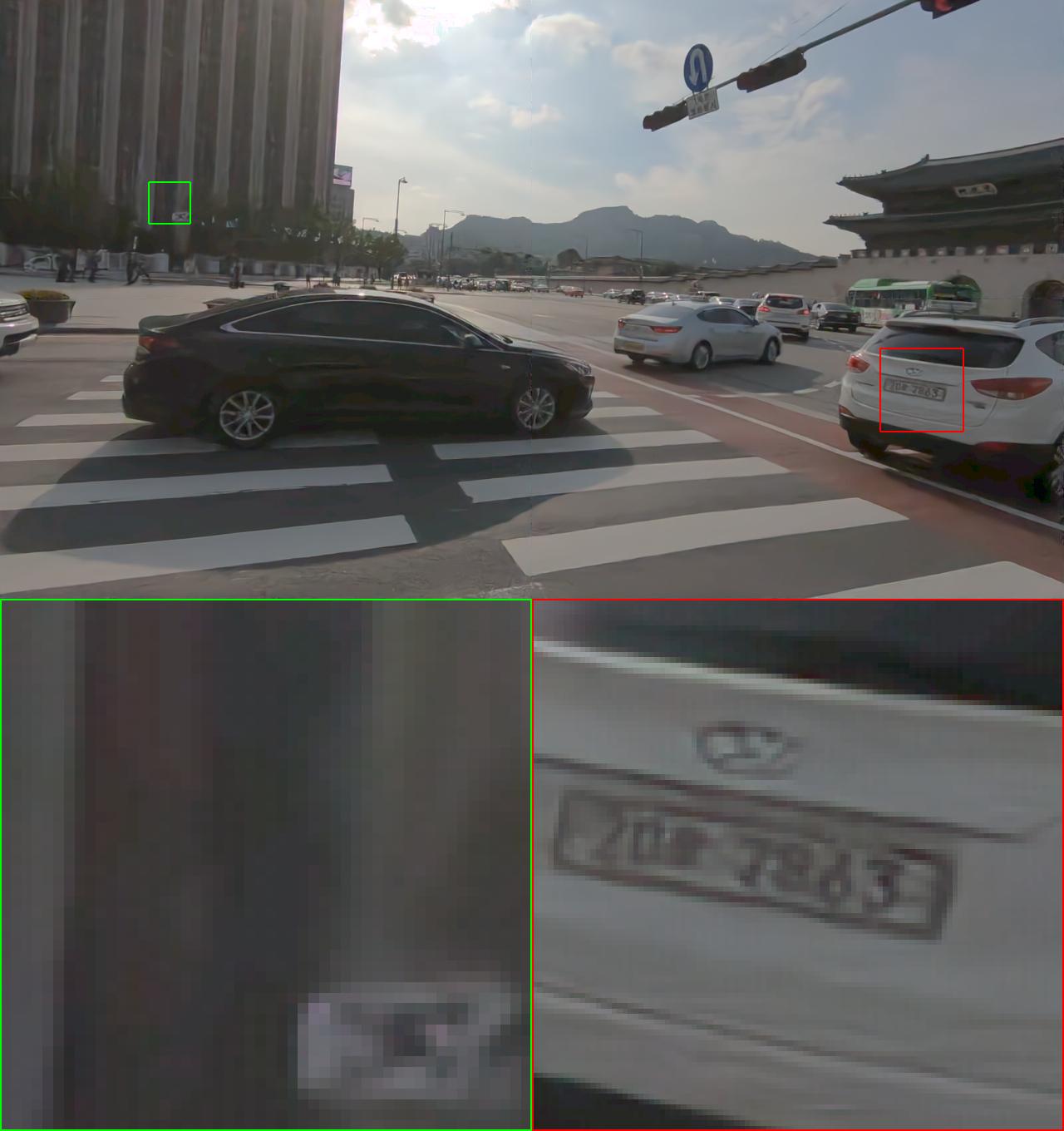}
{\footnotesize  (e) RViDeNet+DMN}
\end{minipage}\hspace{0.05cm}

\begin{minipage}[t]{0.24\textwidth}
\centering
\includegraphics[width=1\textwidth]{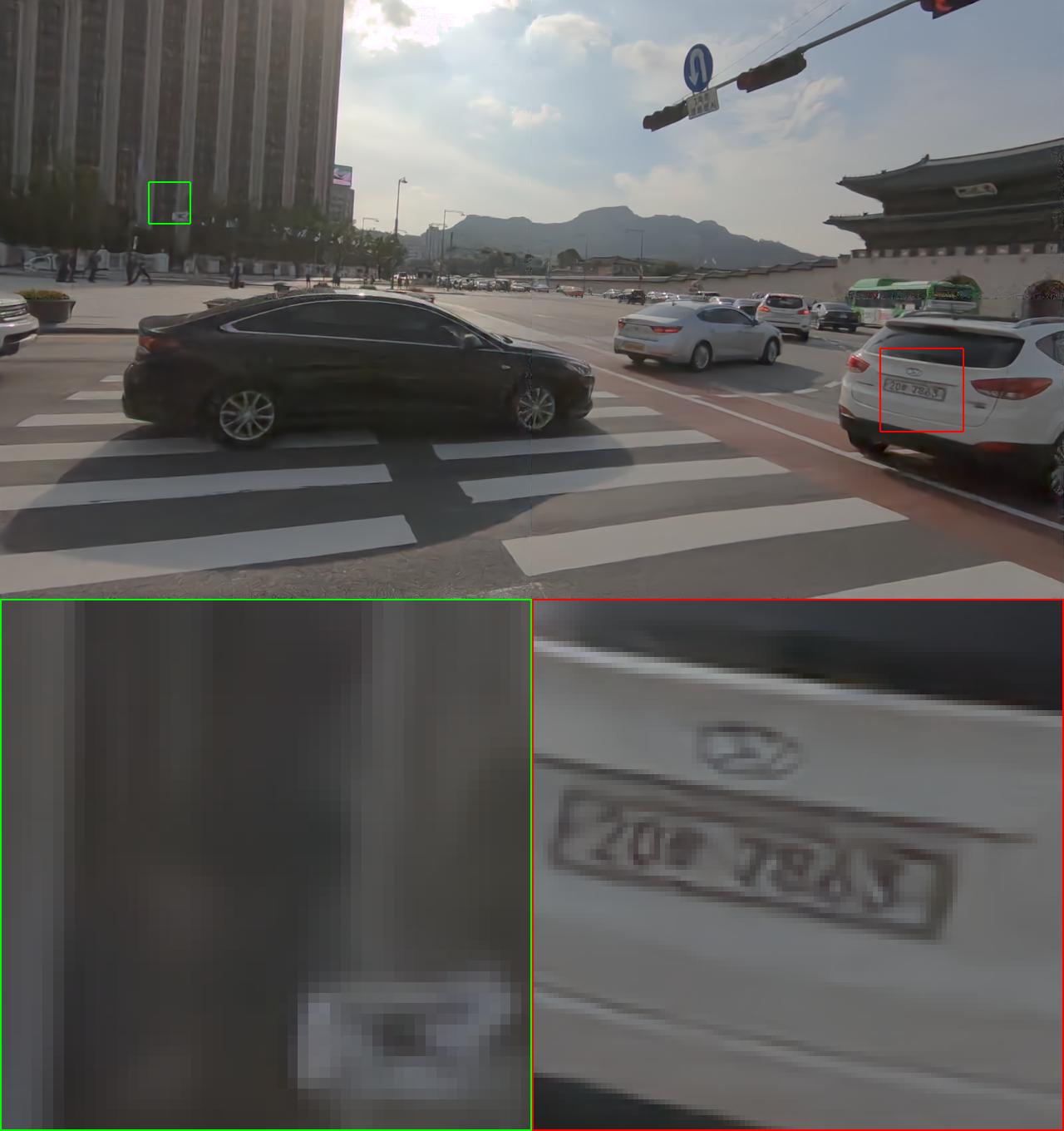}
{\footnotesize  (f) RViDeNet*}
\end{minipage}\hspace{0.05cm}

\begin{minipage}[t]{0.24\textwidth}
\centering
\includegraphics[width=1\textwidth]{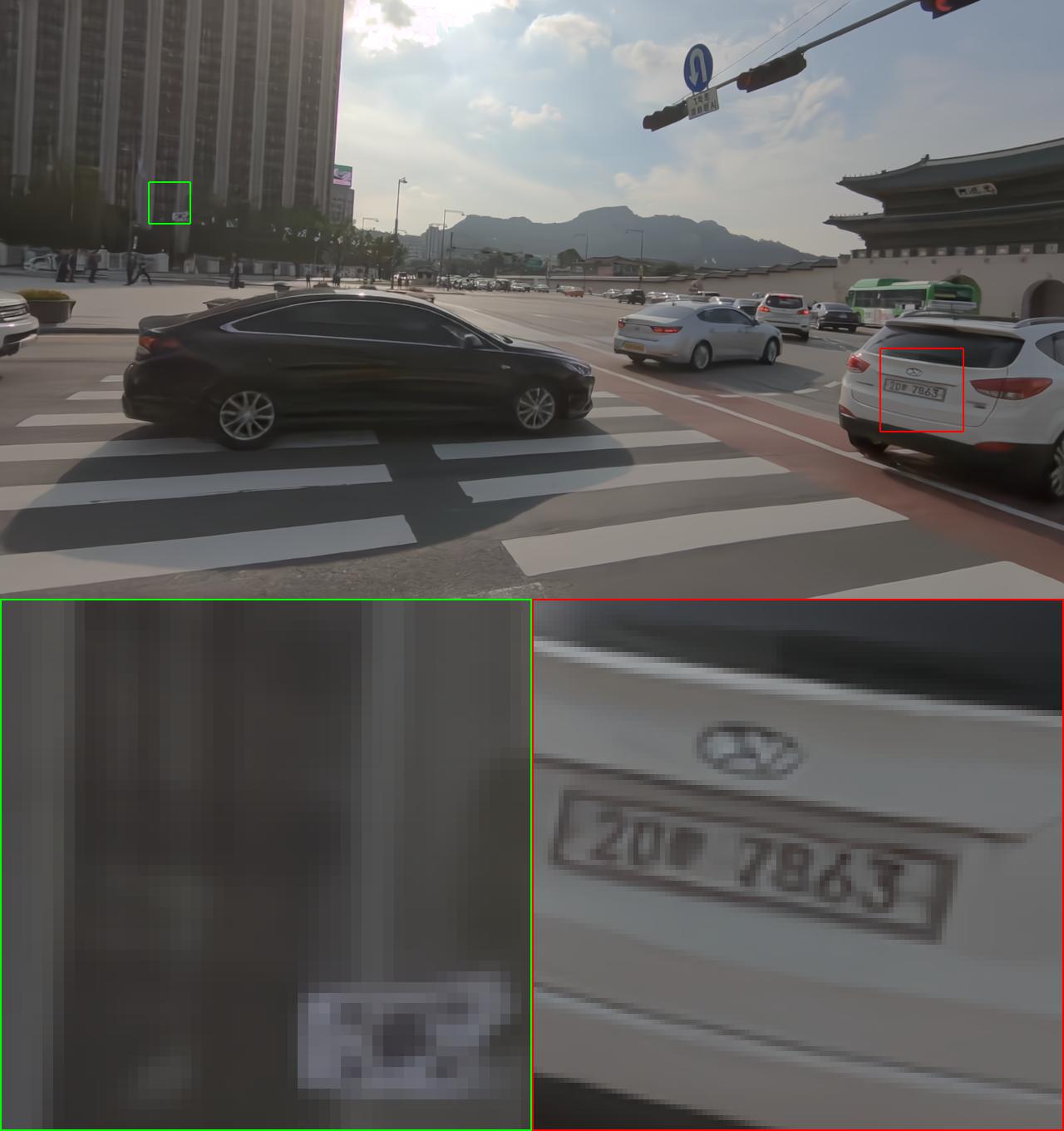}
{\footnotesize  (g) Ours}
\end{minipage}\hspace{0.05cm}

\begin{minipage}[t]{0.24\textwidth}
\centering
\includegraphics[width=1\textwidth]{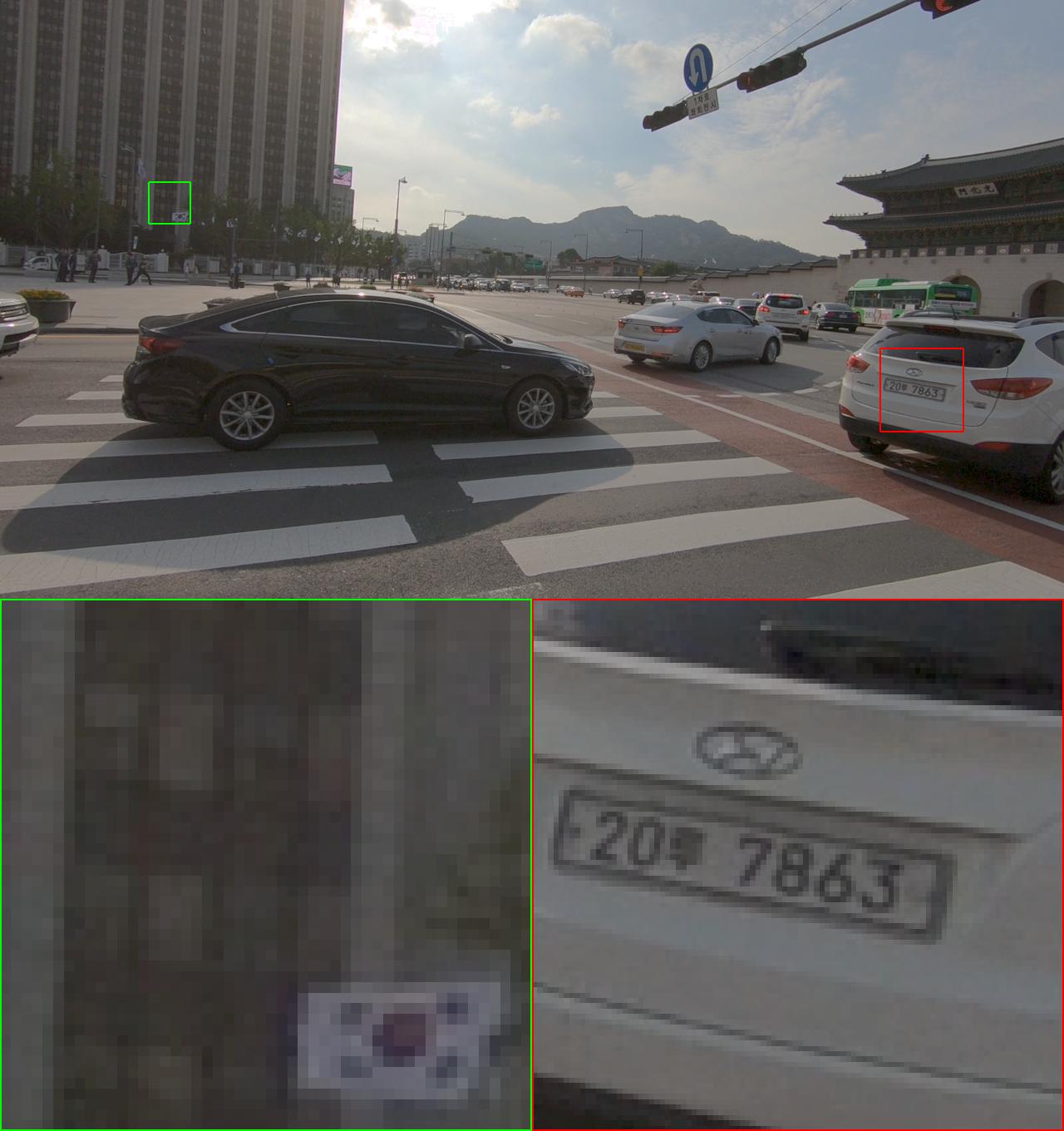}
{\footnotesize  (h) GT}
\end{minipage}
}
\caption{JDD results on clip \emph{015} in REDS4 dataset by different methods.}%\vspace{-0.4cm}
\label{figRED}
\end{figure*}

\subsubsection{Selection of frame number}
In this section, we evaluate the performance of GCP-Net trained with different number of frames, denoted as GCP-Net-$N$ with $N=1,3,5,7$. The results are listed in Tables~\ref{synthVid4} and~\ref{synthREDS4}. Compared with the network using a single image as input, the network using three frames as input can achieve performance gain by a great margin (\emph{i.e.}, 1.1$\sim$1.5dB on Vid4 and 0.3$\sim$0.4dB on REDS4). By using more frames, GCP-Net-5 further improves GCP-Net-3 by about 0.5dB at high noise level and about 0.4dB at low noise level. By further increasing the frame number from 5 to 7, GCP-Net-7 achieves slight improvement ($\sim$0.1dB) on the Vid4 dataset and comparable results on the REDS4 dataset. Considering the computational efficiency and the performance gains, we choose to use $N=5$ frames in our model.

\begin{table*}[!h]
\setlength{\abovecaptionskip}{0.2cm}
\footnotesize
%\small
\centering
%\smallskip
\caption{Quantitative comparison of different JDD-B approaches on the \textbf{Vid4} dataset. Following the experient setting of \cite{mildenhall2018burst,xu2019learning}, ``Low noise level" and ``High noise level" are corresponding to $\sigma_s = 2.5\times 10^{-3}$, $\sigma_r = 10^{-2}$ and $\sigma_s = 6.4\times 10^{-3}$, $\sigma_r = 2\times 10^{-2}$, respectively. The best two results are highlighted in \textcolor{red}{red} and \textcolor{blue}{blue} respectively. The GCP-Net-$N$ represents the network trained with consecutive $N$ frames.}
\label{synthVid4}
%\begin{tabularx}{\linewidth} {c c c c c c c c}
%\begin{tabularx} {\linewidth} { @{}  p{2.5cm}  p{2.0cm}  X X X X X@{}}
\begin{tabularx} {\linewidth} {P{2.5cm} p{2.15cm} P{2.1cm} P{2.1cm} P{2.1cm} P{2.1cm} P{2.1cm}}
\toprule
Noise Level & Methods  &\emph{Calendar} &\emph{City} &\emph{Foliage} &\emph{Walk} &Average \\
\midrule
\multirow{12}{*}{High noise level} 
& FlexISP &20.89/0.6108 &25.61/0.6015 &22.41/0.5908 &23.73/0.5125 &23.16/0.5789 \\
& ADMM  &23.03/0.7187 &29.80/0.7726 &26.06/0.6709 &28.68/0.8459 &26.89/0.7520  \\
& VBM3D+DMN  &23.07/0.7279 &30.22/0.7881 &25.76/0.6344 &28.38/0.8274 &26.86/0.7444  \\
& KPN+DMN  &24.31/0.7989 &30.76/0.8225 &28.09/0.7984 &30.78/0.8818 &28.48/0.8254  \\
& EDVR+DMN  &24.78/0.8228 &31.98/0.8711 &28.19/0.8283 &31.30/0.9001 &29.07/0.8556  \\
& RviDeNet+DMN  &25.82/0.8602 &34.03/0.9153 &30.35/0.8914 &33.20/0.9324 &30.85/0.8998  \\
& EDVR*  &24.59/0.8314 &32.07/0.8670 &28.19/0.8260 &31.78/0.9066 &29.16/0.8578  \\
& RviDeNet*  &\textcolor{red}{27.03/0.8797} &34.40/0.8206 &31.14/0.9065 &34.09/0.9400 &31.67/0.9117  \\
\cmidrule{2-7}
& GCP-Net-1  &26.09/0.8478 &33.08/0.8921 &29.06/0.8471 &32.95/0.9240 &30.29/0.8777  \\
& GCP-Net-3  &26.83/0.8722 &34.68/0.9230 &30.87/0.8971 &34.41/0.9438 &31.70/0.9090  \\
& GCP-Net-5, ours  &26.96/0.8797 &\textcolor{blue}{35.58/0.9367} &\textcolor{blue}{31.70/0.9154} &\textcolor{blue}{34.97/0.9504} & \textcolor{blue}{32.30/0.9205} \\
& GCP-Net-7  &\textcolor{blue}{27.02/0.8865} &\textcolor{red}{35.60/0.9409} &\textcolor{red}{32.05/0.9240} &\textcolor{red}{35.10/0.9529} &\textcolor{red}{32.44/0.9261} \\

%& Ours(w/o LSTM)  &26.97/0.8810 &35.35/0.9341 &31.67/0.9151 &34.89/0.9495 &32.22/0.9199 \\
%& Ours(w RBG)  &26.84/0.8718 &34.74/.9209 &31.22/0.9036 &34.56/0.9468 &31.84/0.9108 \\
%& Ours(w/o GreenG)  &26.95/0.8799 &35.37/0.9341 &31.61/0.9143 &34.89/0.9498 &32.20/0.9196 \\
%& Ours(w/o Inter)  &26.60/0.8668 &34.24/0.9136 &30.54/0.8868 &34.25/0.9438 &31.41/0.9027 \\

\toprule

\multirow{12}{*}{Low noise level} 
& FlexISP &22.28/0.7292 &28.26/0.7692 &24.57/0.7333 &27.17/0.6958 &25.57/0.7319  \\
& ADMM  &23.61/0.7339 &30.50/0.7914 &26.80/0.6944 &30.27/0.8659 &27.79/0.7714  \\
& VBM3D+DMN  &23.90/0.7606 &32.31/0.8558 &26.86/0.6847 &30.13/0.8573 &28.30/0.7896  \\
& KPN+DMN  &25.16/0.8425 &33.09/0.8920 &30.52/0.8868 &32.95/0.9233 &30.43/0.8860  \\
& EDVR+DMN  &25.68/0.8579 &34.17/0.9206 &30.21/0.8931 &33.32/0.9340 &30.85/0.9014  \\
& RviDeNet+DMN  &26.26/0.8775 &35.84/0.9443 &32.23/0.9312 &34.86/0.9536 &32.30/0.9267  \\
& EDVR*  &25.17/0.8630 &34.51/0.9207 &30.25/0.8927 &34.03/0.9416 &30.99/0.9220  \\
& RviDeNet*  &\textcolor{red}{28.03/0.9024} &36.54/0.9501 &33.23/0.9428 &36.31/0.9628 &33.53/0.9395  \\
\cmidrule{2-7}
& GCP-Net-1  &27.16/0.8819 &35.46/0.9347 &31.42/0.9104 &35.38/0.9525 &32.35/0.9199  \\
& GCP-Net-3  &27.58/0.8943 &36.77/0.9517 &32.97/0.9383 &36.52/0.9639 &33.47/0.9371  \\
& GCP-Net-5, ours  &27.58/0.8976 &\textcolor{blue}{37.45/0.9584} &\textcolor{blue}{33.67/0.9479} &\textcolor{blue}{36.96/0.9677} &\textcolor{blue}{33.91/0.9429}  \\
& GCP-Net-7 &\textcolor{blue}{27.66/0.9010} &\textcolor{red}{37.59/0.9607} &\textcolor{red}{33.86/0.9520} &\textcolor{red}{37.00/0.9689} &\textcolor{red}{34.02/0.9456}\\

%& Ours(w/o LSTM)  &27.69/0.8993 &37.34/0.9575 &33.67/0.9460 &36.90/0.9673 &33.90/0.9430  \\
%& Ours(w RBG)  &27.70/0.8963 &36.85/0.9515 &33.30/0.9428 &36.63/0.9658 &33.62/0.9391 \\
%& Ours(w/o GreenG)  &27.67/0.8978 &37.30/0.9575 &33.55/0.9468 &36.86/0.9672 &33.84/0.9423 \\
%& Ours(w/o Inter)  &27.43/0.8919 &36.34/0.9453 &32.63/0.9307 &36.39/0.9635 &33.20/0.9329 \\

\toprule

\end{tabularx} 
\end{table*}

\begin{table*}[!tbp]
\setlength{\abovecaptionskip}{0.2cm}
\footnotesize
%\small
\centering
%\smallskip
\caption{Quantitative comparison of different JDD-B approaches on the \textbf{REDS4} dataset. Following the experient setting of \cite{mildenhall2018burst,xu2019learning}, ``Low noise level" and ``High noise level" are corresponding to $\sigma_s = 2.5\times 10^{-3}$, $\sigma_r = 10^{-2}$ and $\sigma_s = 6.4\times 10^{-3}$, $\sigma_r = 2\times 10^{-2}$, respectively. The best two results are highlighted in \textcolor{red}{red} and \textcolor{blue}{blue} respectively. The GCP-Net-$N$ represents the network trained with consecutive $N$ frames.}
\label{synthREDS4}
%\begin{tabularx}{\linewidth} {c c c c c c c c}
%\begin{tabularx} {\linewidth} { @{}  p{2.5cm}  p{2.0cm}  X X X X X@{}}
\begin{tabularx} {\linewidth} {P{2.5cm} p{2.15cm} P{2.1cm} P{2.1cm} P{2.1cm} P{2.1cm} P{2.1cm}}
\toprule
Noise Level & Methods &\emph{Clip000} &\emph{Clip011} &\emph{Clip015} &\emph{Clip020} &Average \\
\midrule
\multirow{12}{*}{High noise level} 
& FlexISP &23.44/0.5257 &24.09/0.4820 &24.37/0.4338 &23.75/0.5129 &23.91/0.4886  \\
& ADMM  &26.35/0.6897 &28.97/0.7928 &29.08/0.8204 &27.74/0.7919 &28.04/0.7737  \\
& VBM3D+DMN  &26.04/0.6672 &27.87/0.7536 &28.30/0.7708 &26.52/0.7432 &27.19/0.7337  \\
& KPN+DMN  &27.36/0.7211 &29.22/0.7703 &31.14/0.8299 &28.36/0.7798 &29.02/0.7753  \\
& EDVR+DMN  &29.20/0.8164 &30.76/0.8314 &33.15/0.8966 &30.19/0.8498 &30.83/0.8443  \\
& RviDeNet+DMN  &30.28/0.8519 &32.06/0.8553 &33.91/0.8883 &31.30/0.8737 &31.89/0.8673  \\
& EDVR*  &29.33/0.8279 &31.67/0.8535 &33.47/0.8870 &30.99/0.8731 &31.37/0.8607  \\
& RviDeNet*  &31.35/0.8816 &32.57/0.8672 &34.88/0.9064 &31.87/0.8853 &32.67/0.8851  \\
\cmidrule{2-7}
& GCP-Net-1  &30.30/0.8557 &32.94/0.8729 &34.91/0.9069 &32.32/0.8926 &32.62/0.8820  \\
& GCP-Net-3  &31.46/0.8818 &32.97/0.8731 &35.37/0.9115 &32.30/0.8926 &33.03/0.8898  \\
& GCP-Net-5, ours  &\textcolor{blue}{32.29/0.9035} &\textcolor{red}{33.28/0.8783} &\textcolor{blue}{35.79/0.9177} &\textcolor{red}{32.63/0.8979} &\textcolor{blue}{33.50/0.8994} \\
& GCP-Net-7  &\textcolor{red}{32.39/0.9044} &\textcolor{blue}{33.21/0.8773} &\textcolor{red}{35.86/0.9180} &\textcolor{blue}{32.58/0.8975} &\textcolor{red}{33.51/0.8993} \\

%& Ours(w/o LSTM)  &32.14/0.8995 &33.24/0.8784 &35.79/0.9179 &32.61/0.8977 &33.44/0.8983 \\
%& Ours(w/o GreenG)  &32.02/0.8976 &33.14/0.8769/ &35.64/0.9159 &32.48/0.8958 & 33.32/0.8966 \\
%& Ours(w/o Inter)  &30.76/0.8645 &32.78/0.8712 &35.15/0.9090 &32.05/0.8898 & 32.69/0.8836 \\
%& Ours(w RBG)  &30.98/0.8645 &32.79/0.8697 &35.18/0.9085 &32.11/0.8890 &32.77/0.8829 \\

\toprule

\multirow{12}{*}{Low noise level} 
& FlexISP &25.86/0.6932 &27.49/0.6638 &27.85/0.6314 &26.70/0.6878 &26.97/0.6690  \\
& ADMM  &27.38/0.7139 &30.43/0.8123 &31.17/0.8463 &29.24/0.8143 &29.56/0.7967 \\
& VBM3D+DMN  &27.68/0.7308 &30.17/0.8061 &31.36/0.8402 &28.78/0.8008 &29.50/0.7945 \\
& KPN+DMN  &29.36/0.8078 &31.36/0.8270 &33.23/0.8717 &30.83/0.8480 &31.20/0.8386  \\
& EDVR+DMN  &29.95/0.8484 &31.59/0.8539 &34.05/0.8959 &31.16/0.8756 &31.69/0.8684  \\
& RviDeNet+DMN  &32.20/0.9034 &33.97/0.8932 &35.58/0.9134 &33.45/0.9121 &33.80/0.9056  \\
& EDVR*  &31.44/0.8922 &33.59/0.8917 &35.54/0.9169 &33.16/0.9129 &33.43/0.9034  \\
& RviDeNet*  &33.58/0.9285 &34.63/0.9044 &36.98/0.9327 &34.23/0.9230 &34.86/0.9221  \\
\cmidrule{2-7}
& GCP-Net-1  &32.66/0.9129 &34.99/0.9092 &36.99/0.9329 &34.67/0.9286 &34.82/0.9209 \\
& GCP-Net-3  &33.66/0.9287 &34.95/0.9087 &37.33/0.9351 &34.54/0.9274 &35.12/0.9250  \\
& GCP-Net-5, ours  &\textcolor{blue}{34.45/0.9414} &\textcolor{red}{35.27/0.9144} &\textcolor{blue}{37.74/0.9403} &\textcolor{red}{34.82/0.9311} &\textcolor{red}{35.57/0.9318} \\
& GCP-Net-7 &\textcolor{red}{34.48/0.9415} &\textcolor{blue}{35.14/0.9116} &\textcolor{red}{37.77/0.9397} &\textcolor{blue}{34.77/0.9309} &\textcolor{blue}{35.54/0.9309}  \\

%& Ours(w/o LSTM)  &34.30/0.9393 &35.24/0.9157 &37.75/0.9052 &34.83/0.9310 &35.54/0.9316 \\
%& Ours(w/o GreenG)  &34.15/0.9373 &35.12/0.9131 &37.61/0.9388 &34.69/0.9298 & 35.38/0.9297 \\
%& Ours(w/o Inter)  &32.95/0.9166 &34.81/0.9088 &37.13/0.9336 &34.31/0.9261 & 34.80/0.9213 \\
%& Ours(w RBG)  &33.31/0.9222 &34.89/0.9092 &37.23/0.9343 &34.41/0.9266 &34.96/0.9231 \\
\toprule

\end{tabularx} 
\end{table*}

\subsection{Comparison Methods}
Since currently there is no JDD-B method publically available, for fair comparison, we combine the representative burst image denoising methods with a state-of-the-art demosaicking method, DemosaicNet (DMN)~\cite{gharbi2016deep}, to compare with our GCP-Net. We choose four widely used burst denoising algorithms: VBM3D~\cite{kostadin2007video}, KPN~\cite{mildenhall2018burst}, EDVR~\cite{wang2019edvr} and RViDeNet~\cite{yue2020supervised}. For VBM3D, the noise level is needed as the input, and we use the method in \cite{chen2015efficient} for noise level estimation. For the KPN, EDVR and RViDeNet models, they are retrained using the same training data as GCP-Net. The retrained EDVR and RViDeNet models adopt 20 residual blocks to perform feature extraction and 40 residual blocks in the reconstruction module. The size of learned per-pixel kernel of retrained KPN is 7 x 7.

We also adjust the structures of EDVR and RViDeNet by adding one upsampling operator in the reconstruction step, and train these models for the JDD-B task. These models are denoted as EDVR* and RViDeNet*, respectively. In addition, we compare with two state-of-the-art JDD-S algorithms: FlexISP~\cite{heide2014flexisp} and ADMM~\cite{tan2017joint}.

%\label{synvideo}
\begin{figure*}[!t]
\setlength{\abovecaptionskip}{0.0cm}
\setlength{\belowcaptionskip}{-0.cm}
\centering

\begin{minipage}[b]{1.0\textwidth}
\centering
    \begin{minipage}[b]{0.24\textwidth}
    \centering
    \includegraphics[width=1\textwidth]{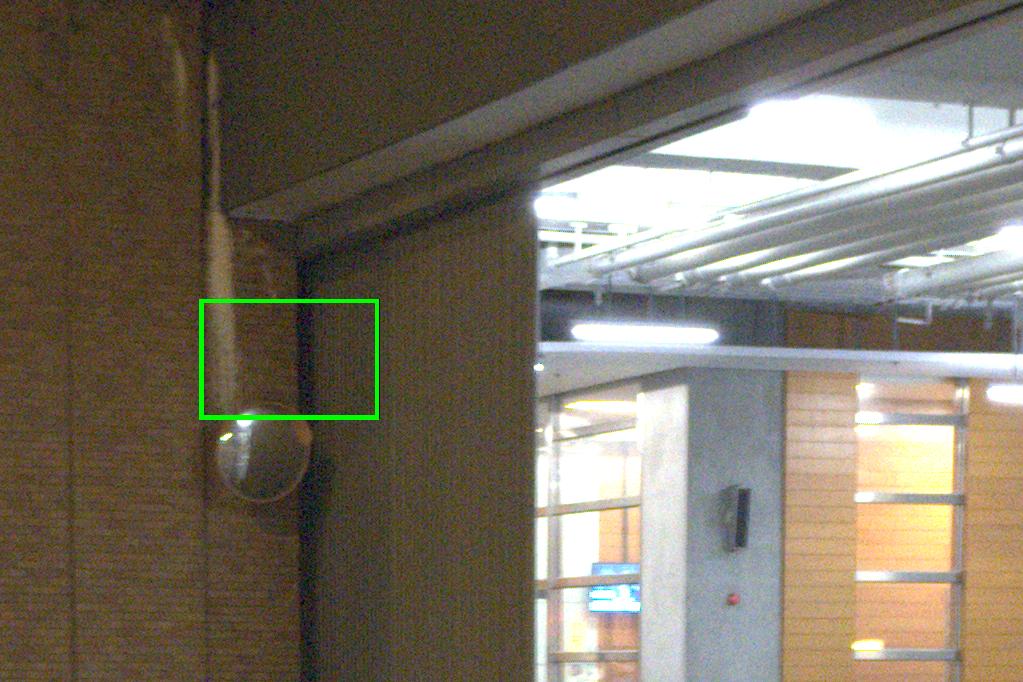}
    {\small Noise image (iPhone X)}
    \end{minipage} %\hspace{0.1cm}
    \begin{minipage}[b]{0.24\textwidth}
    \centering
    \includegraphics[width=1\textwidth]{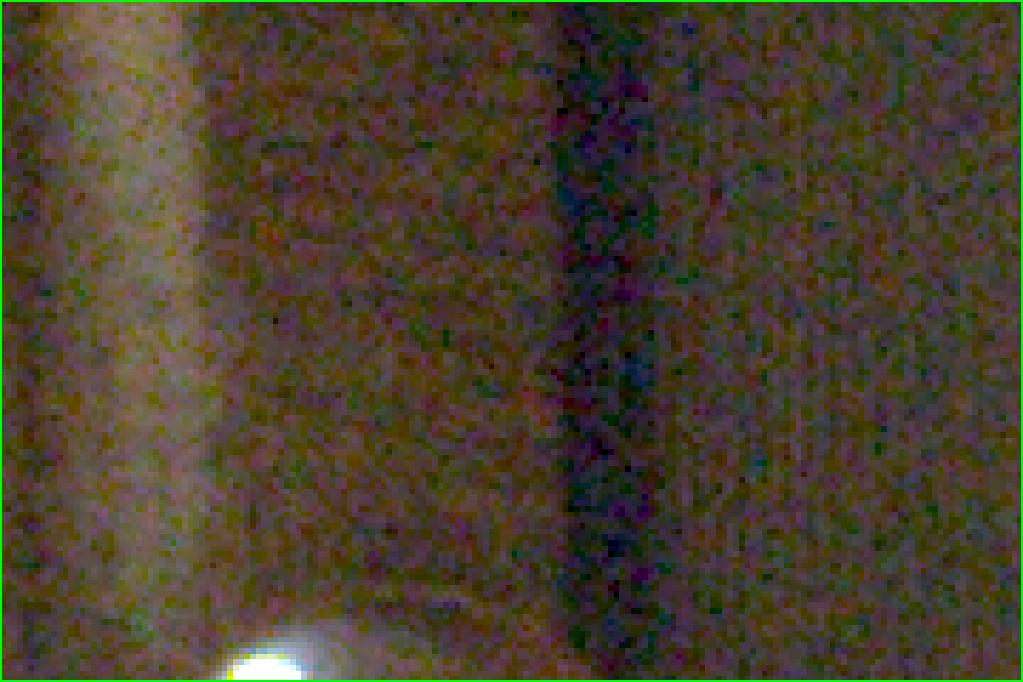}
    {\small Noise patch}
    \end{minipage} %\hspace{0.1cm}
    \begin{minipage}[b]{0.24\textwidth}
    \centering
    \includegraphics[width=1\textwidth]{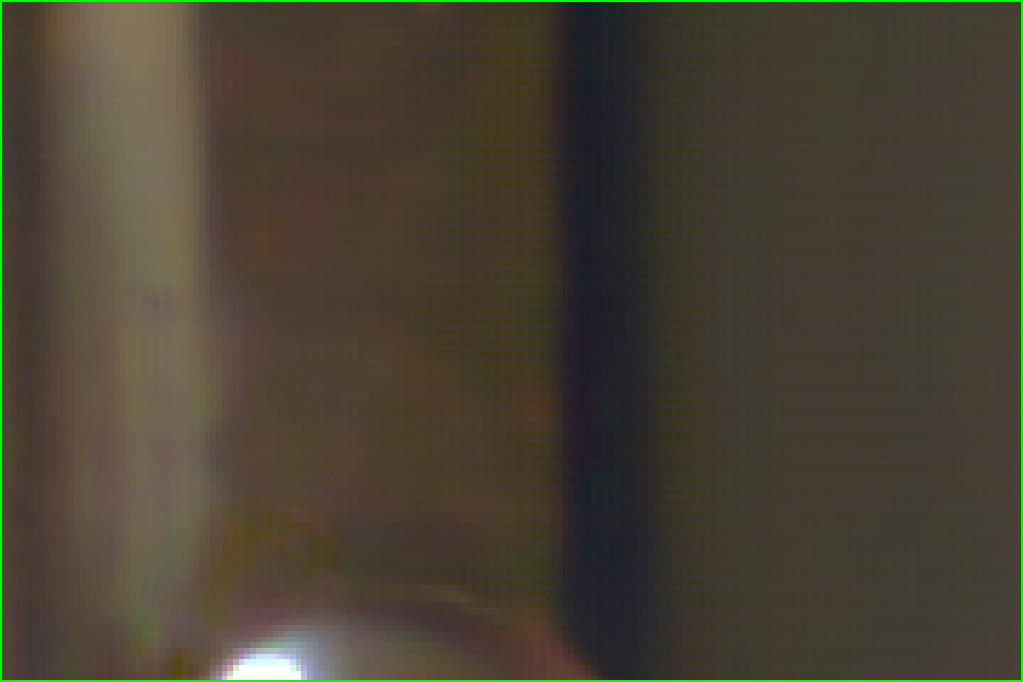}
    {\small KPN+DMN}
    \end{minipage} 
    \begin{minipage}[b]{0.24\textwidth}
    \centering
    \includegraphics[width=1\textwidth]{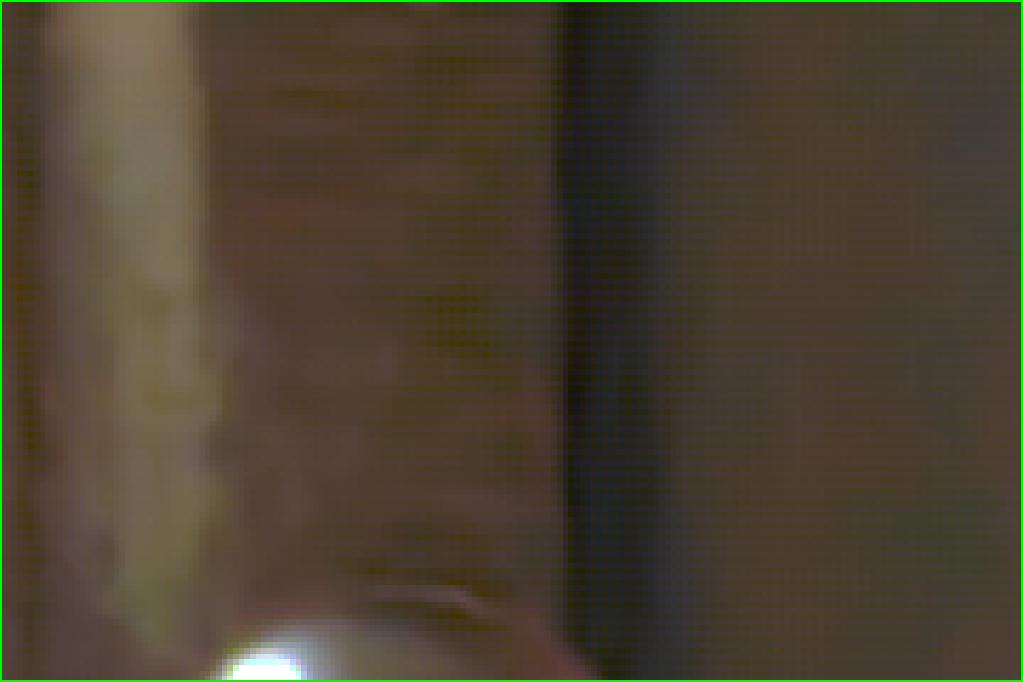}
    {\small EDVR+DMN}
    \end{minipage} %\vspace{0.10cm}
\end{minipage}%\hspace{0.1cm}
\vspace{0.15cm}

\begin{minipage}[b]{1.0\textwidth}
\centering
    \begin{minipage}[b]{0.24\textwidth}
    \centering
    \includegraphics[width=1\textwidth]{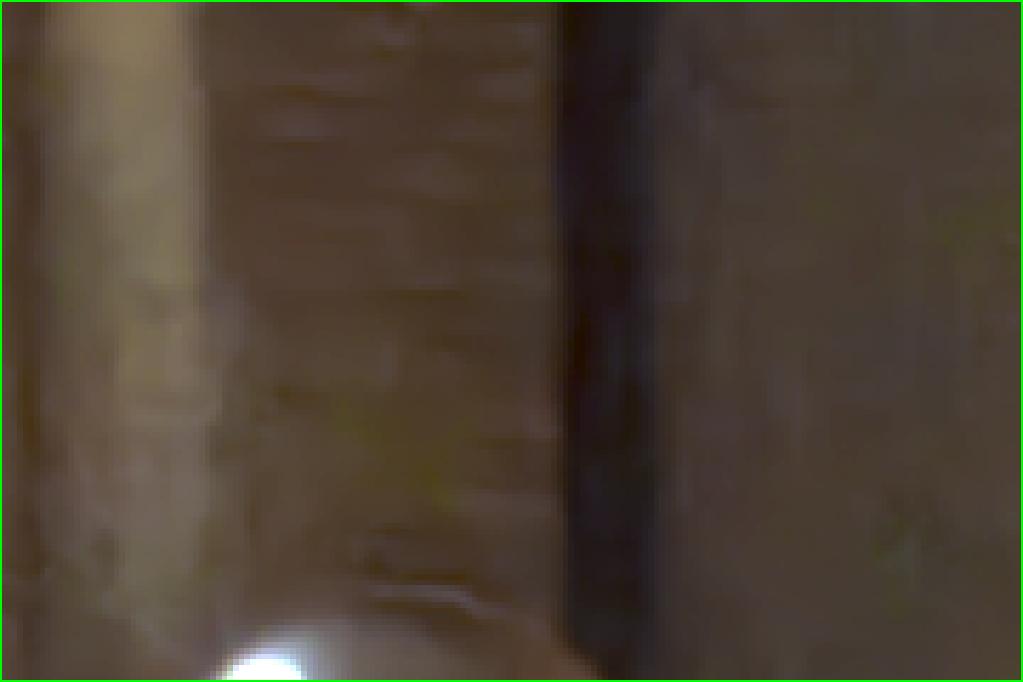}
    {\small EDVR*}
    \end{minipage} %\hspace{0.1cm}
    \begin{minipage}[b]{0.24\textwidth}
    \centering
    \includegraphics[width=1\textwidth]{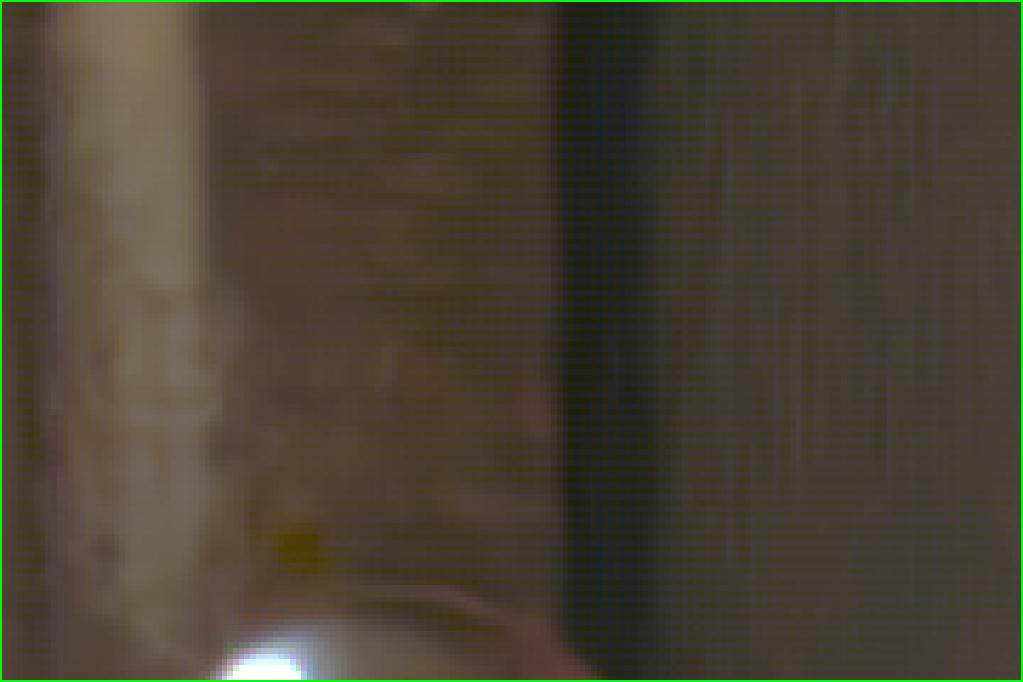}
    {\small RviDeNet+DMN}
    \end{minipage} %\hspace{0.1cm}
    \begin{minipage}[b]{0.24\textwidth}
    \centering
    \includegraphics[width=1\textwidth]{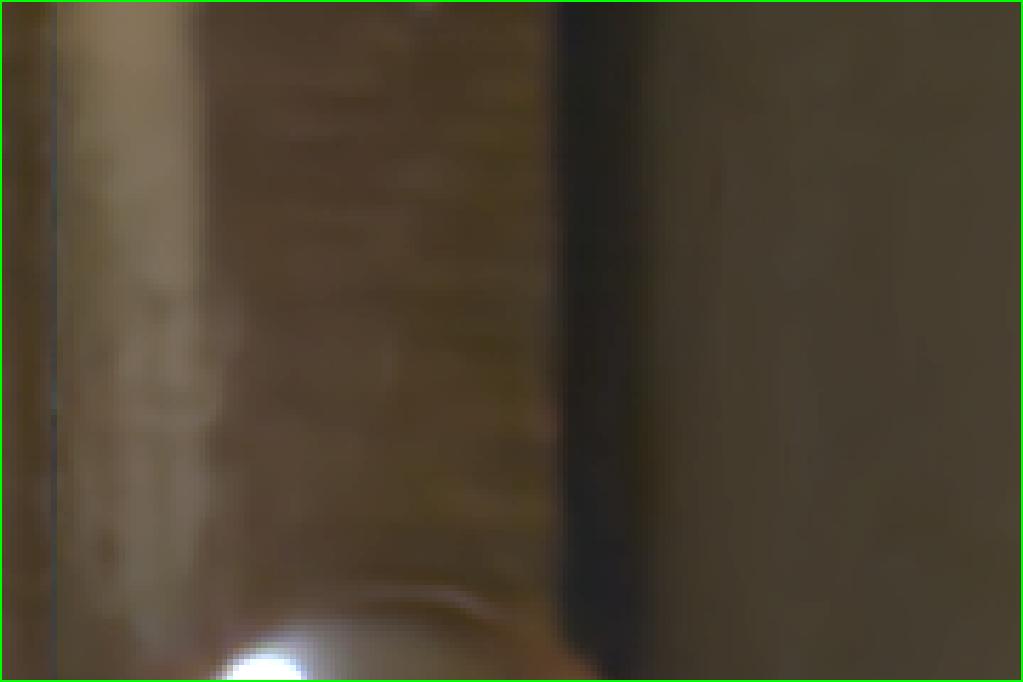}
    {\small RviDeNet*}
    \end{minipage} 
    \begin{minipage}[b]{0.24\textwidth}
    \centering
    \includegraphics[width=1\textwidth]{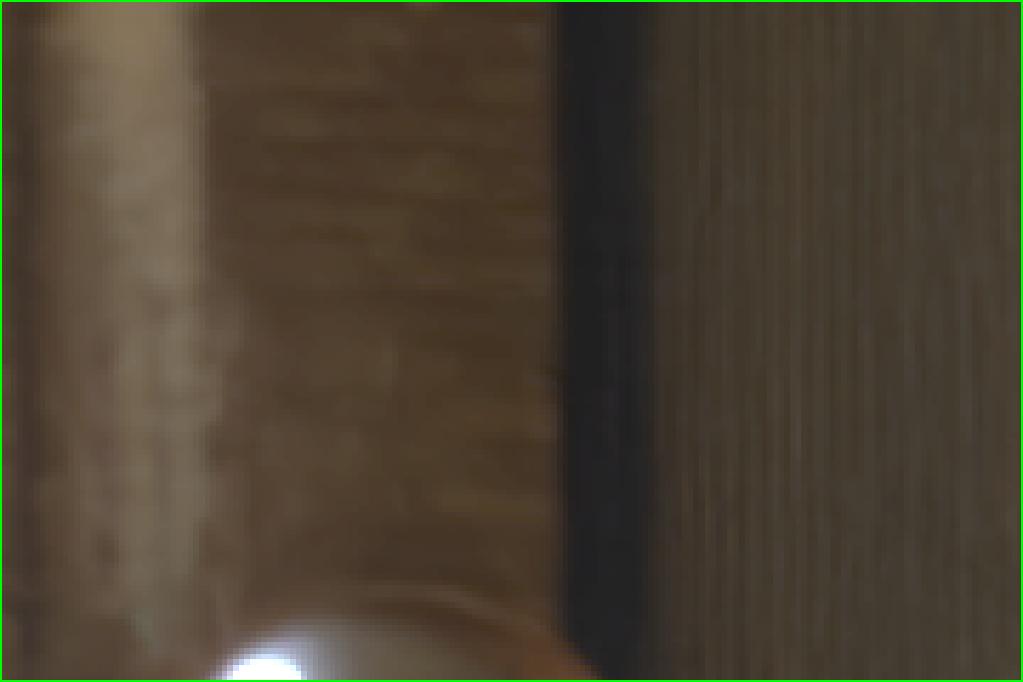}
    {\small Ours}
    \end{minipage} 
\end{minipage}%\hspace{0.1cm}
\vspace{0.15cm}

%%%% patch image 2
\begin{minipage}[b]{1.0\textwidth}
\centering
    \begin{minipage}[b]{0.24\textwidth}
    \centering
    \includegraphics[width=1\textwidth]{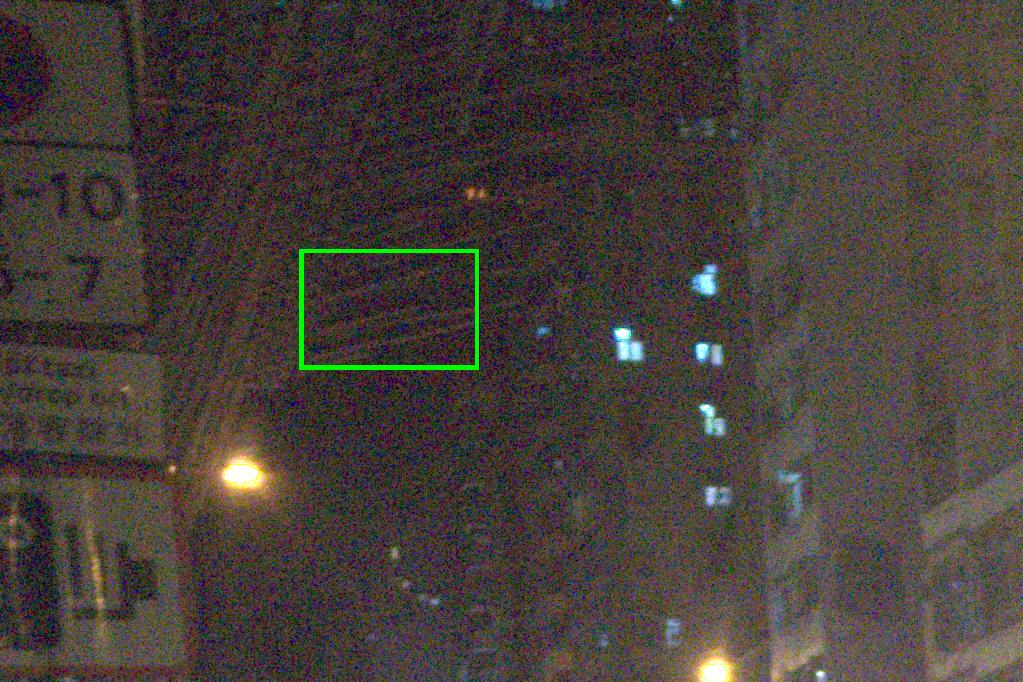}
    {\small Noise image (Pixel 2)}
    \end{minipage} %\hspace{0.1cm}
    \begin{minipage}[b]{0.24\textwidth}
    \centering
    \includegraphics[width=1\textwidth]{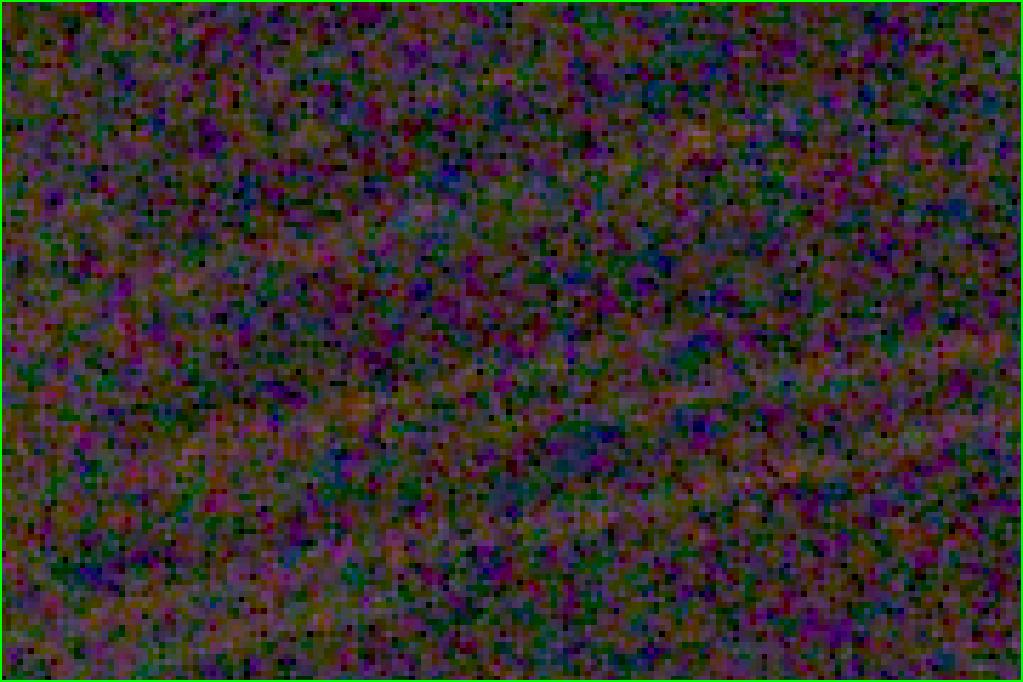}
    {\small Noise patch}
    \end{minipage} %\hspace{0.1cm}
    \begin{minipage}[b]{0.24\textwidth}
    \centering
    \includegraphics[width=1\textwidth]{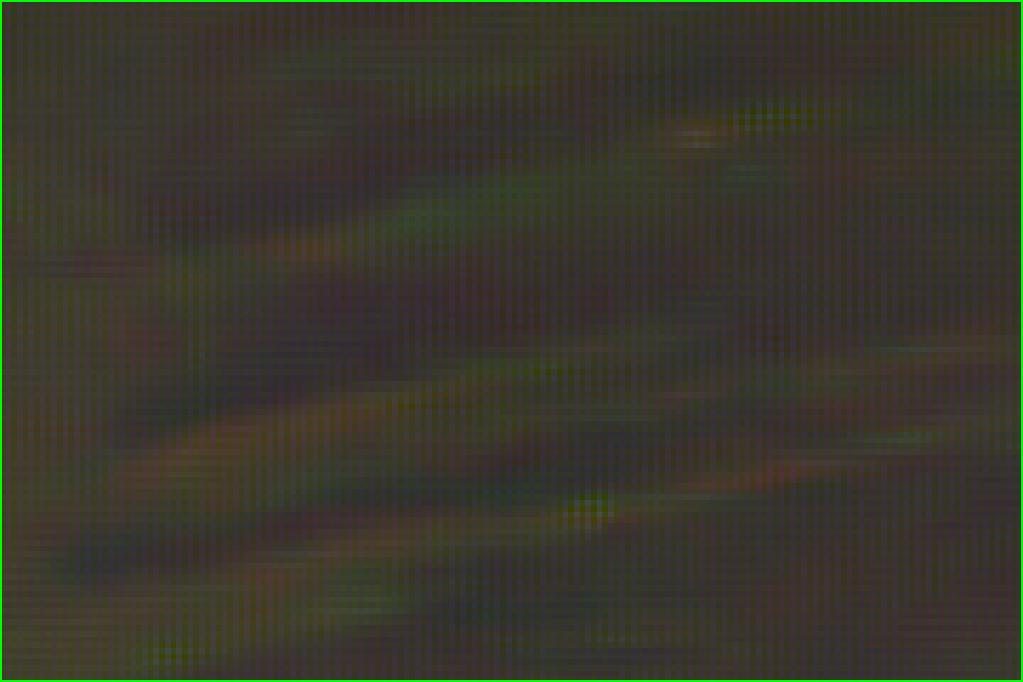}
    {\small KPN+DMN}
    \end{minipage} 
    \begin{minipage}[b]{0.24\textwidth}
    \centering
    \includegraphics[width=1\textwidth]{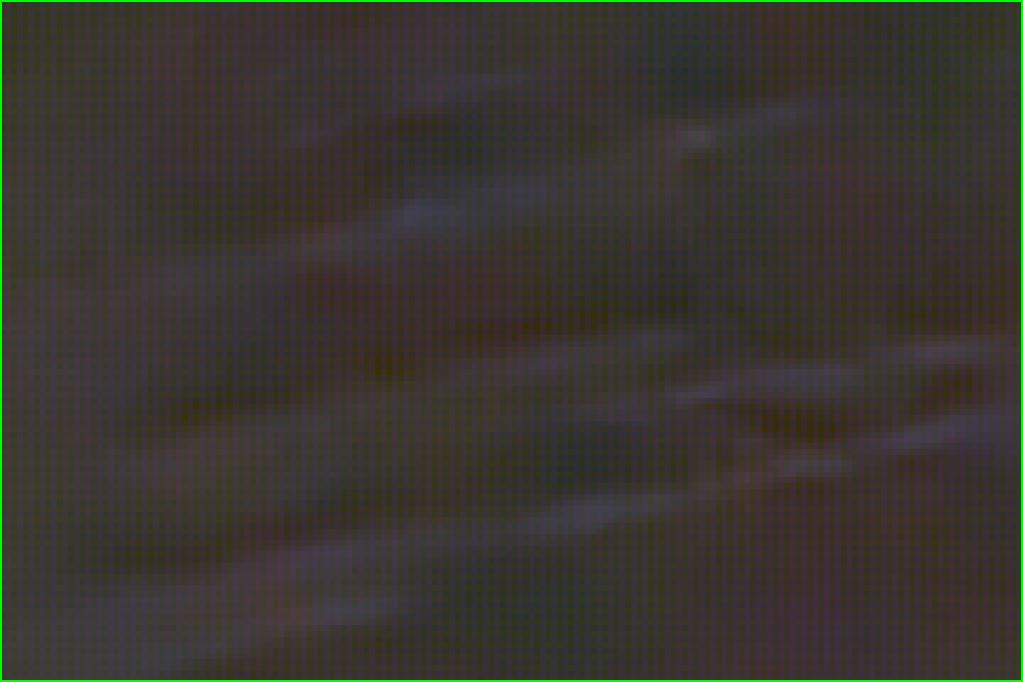}
    {\small EDVR+DMN}
    \end{minipage} %\vspace{0.10cm}
\end{minipage}%\hspace{0.1cm}
\vspace{0.15cm}

\begin{minipage}[b]{1.0\textwidth}
\centering
    \begin{minipage}[b]{0.24\textwidth}
    \centering
    \includegraphics[width=1\textwidth]{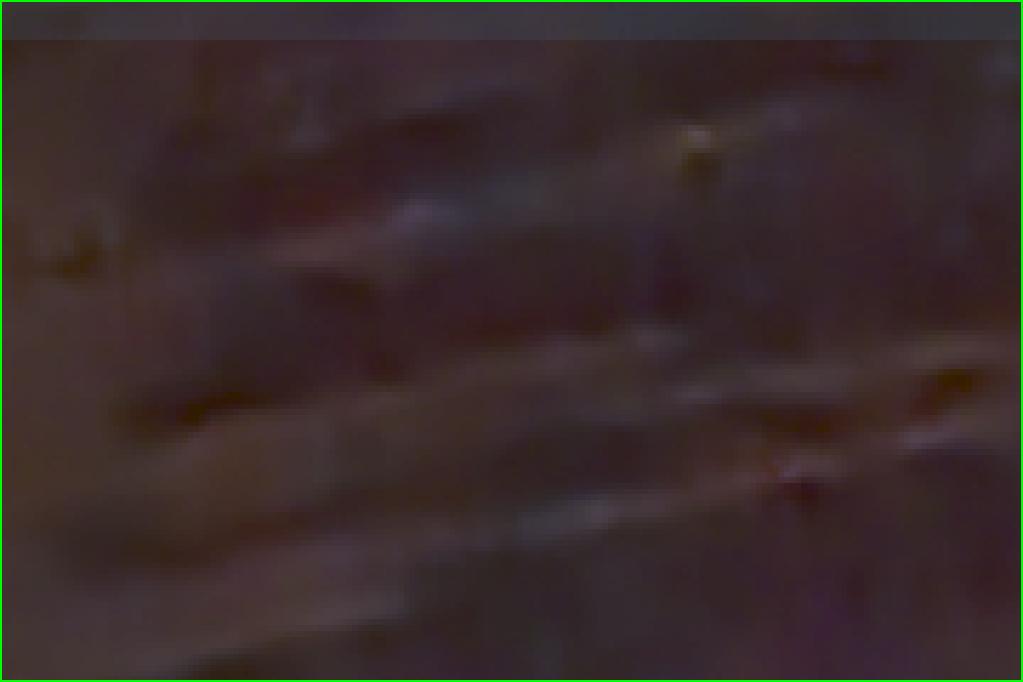}
    {\small EDVR*}
    \end{minipage} %\hspace{0.1cm}
    \begin{minipage}[b]{0.24\textwidth}
    \centering
    \includegraphics[width=1\textwidth]{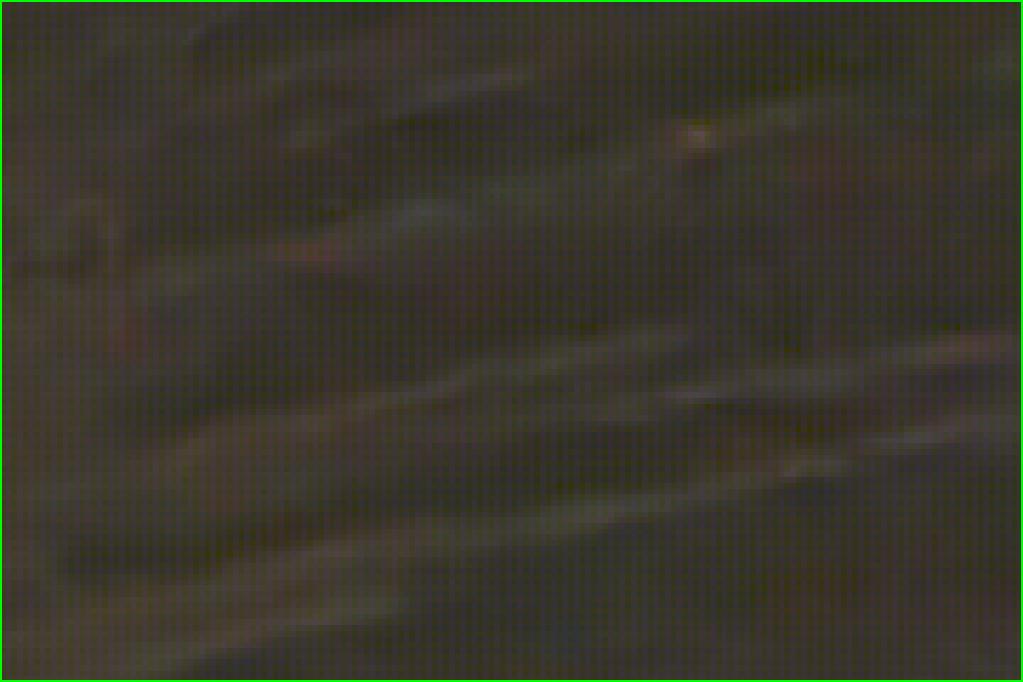}
    {\small RviDeNet+DMN}
    \end{minipage} %\hspace{0.1cm}
    \begin{minipage}[b]{0.24\textwidth}
    \centering
    \includegraphics[width=1\textwidth]{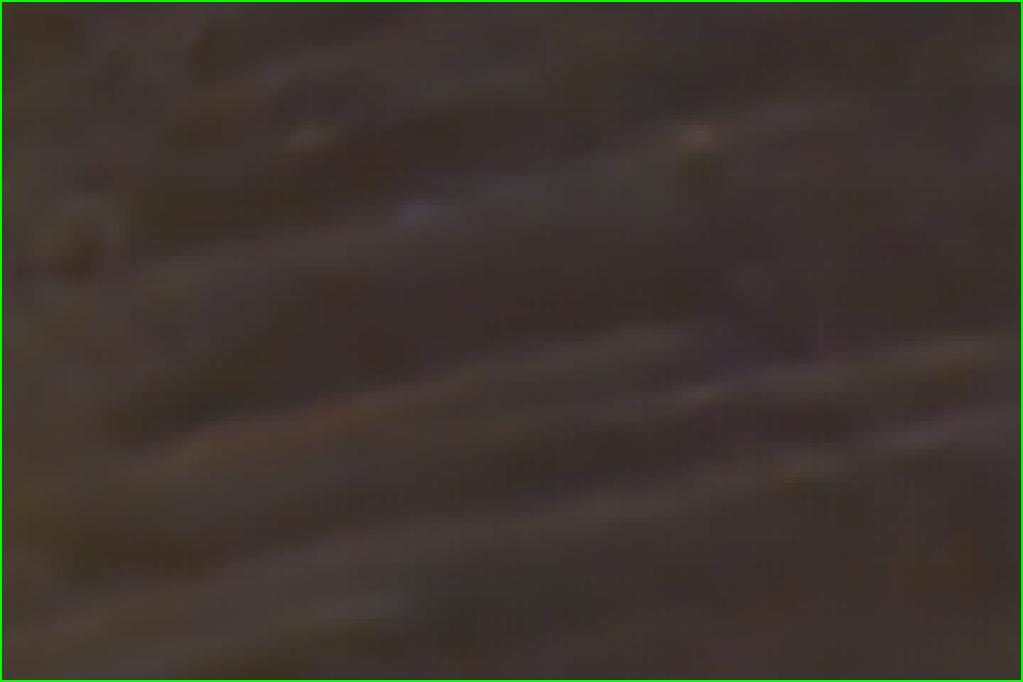}
    {\small RviDeNet*}
    \end{minipage} 
    \begin{minipage}[b]{0.24\textwidth}
    \centering
    \includegraphics[width=1\textwidth]{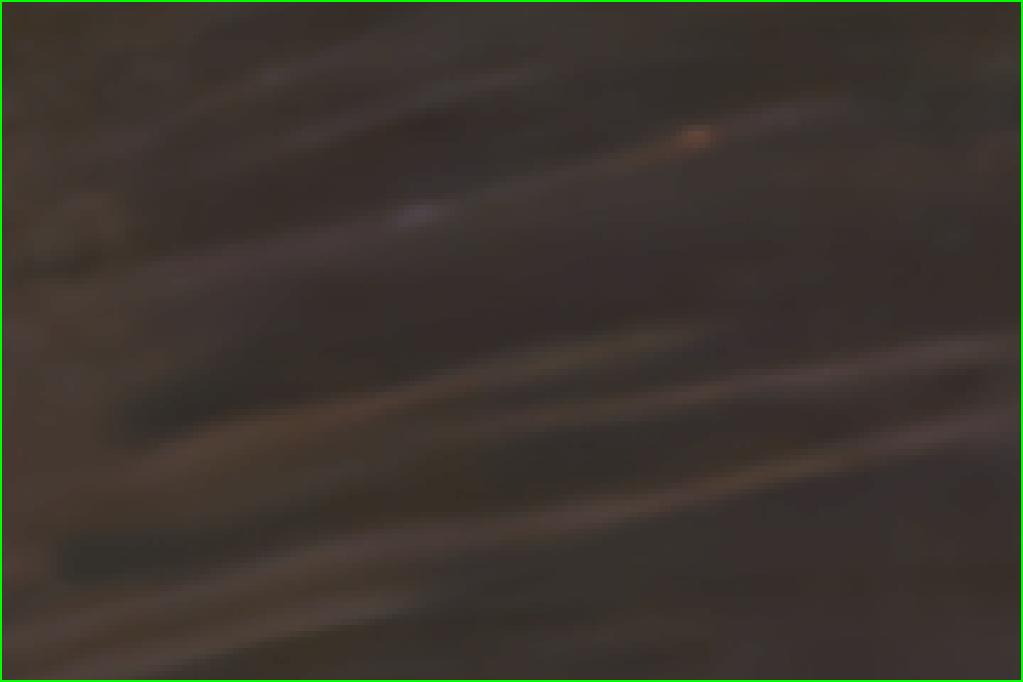}
    {\small Ours}
    \end{minipage} 
\end{minipage}%\hspace{0.1cm}
\vspace{0.15cm}

%%%% patch image 3
\begin{minipage}[b]{1.0\textwidth}
\centering
    \begin{minipage}[b]{0.24\textwidth}
    \centering
    \includegraphics[width=1\textwidth]{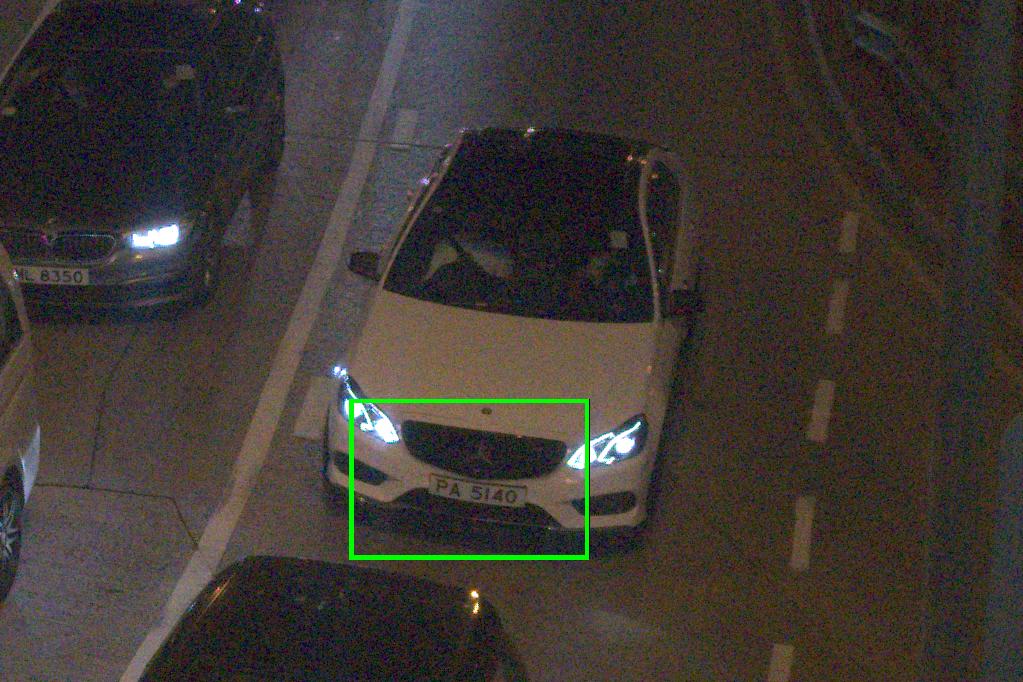}
    {\small Noise image (iPhone 7)}
    \end{minipage} %\hspace{0.1cm}
    \begin{minipage}[b]{0.24\textwidth}
    \centering
    \includegraphics[width=1\textwidth]{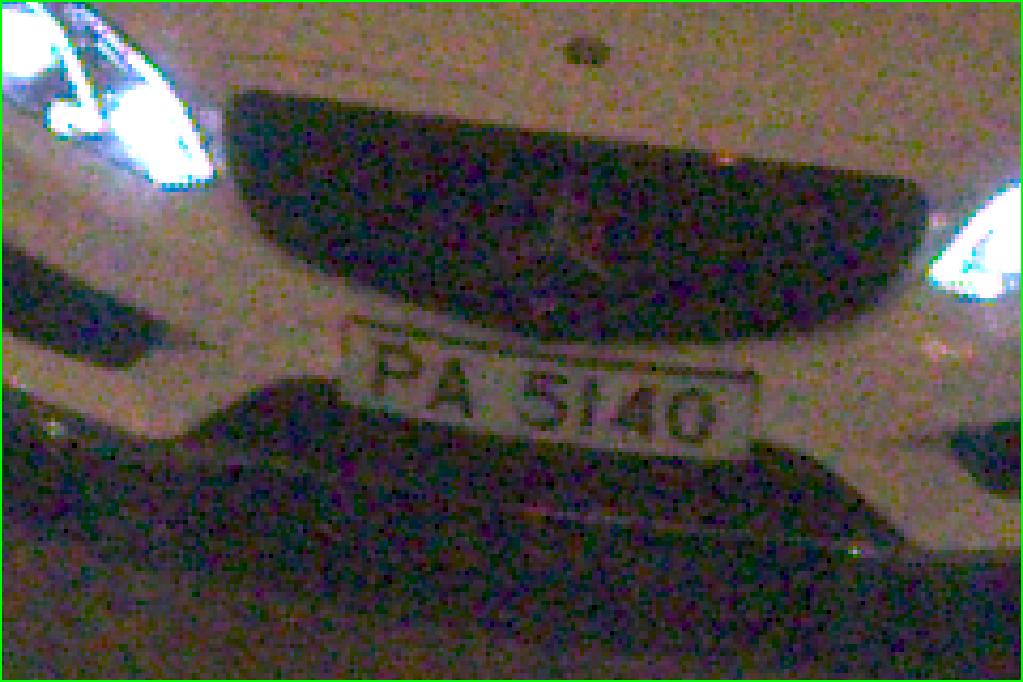}
    {\small Noise patch}
    \end{minipage} %\hspace{0.1cm}
    \begin{minipage}[b]{0.24\textwidth}
    \centering
    \includegraphics[width=1\textwidth]{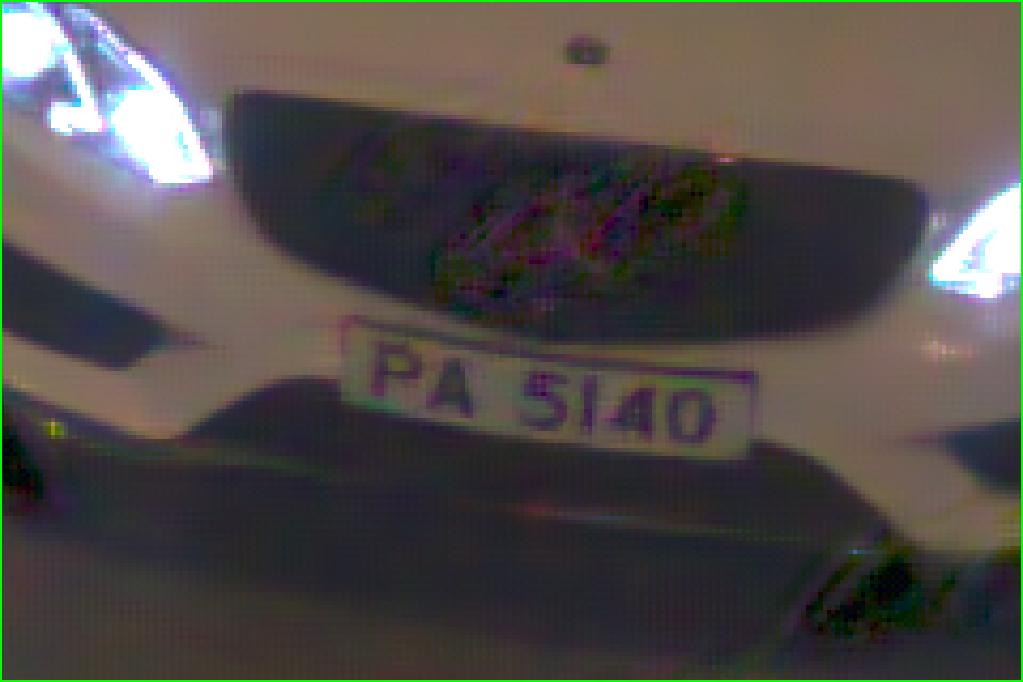}
    {\small KPN+DMN}
    \end{minipage} 
    \begin{minipage}[b]{0.24\textwidth}
    \centering
    \includegraphics[width=1\textwidth]{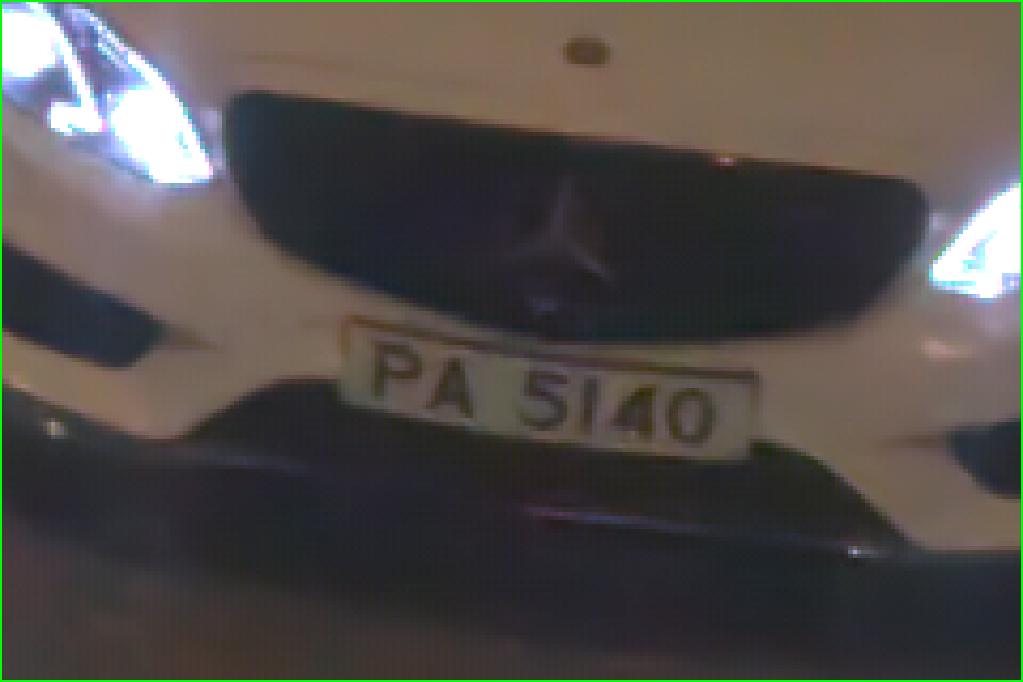}
    {\small EDVR+DMN}
    \end{minipage} %\vspace{0.10cm}
\end{minipage}%\hspace{0.1cm}
\vspace{0.15cm}

\begin{minipage}[b]{1.0\textwidth}
\centering
    \begin{minipage}[b]{0.24\textwidth}
    \centering
    \includegraphics[width=1\textwidth]{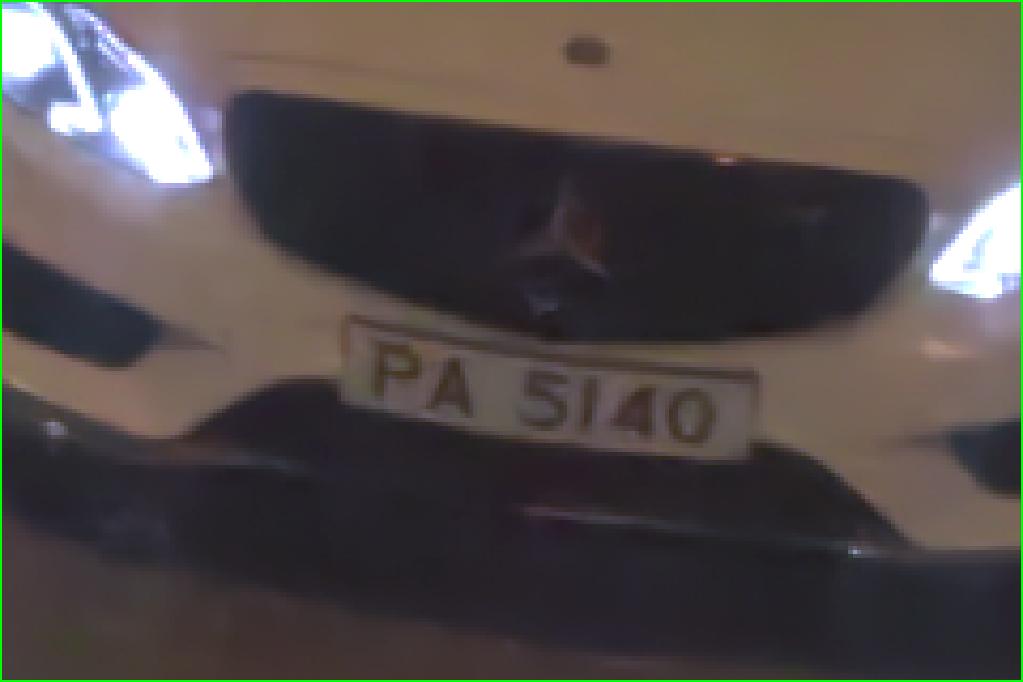}
    {\small EDVR*}
    \end{minipage} %\hspace{0.1cm}
    \begin{minipage}[b]{0.24\textwidth}
    \centering
    \includegraphics[width=1\textwidth]{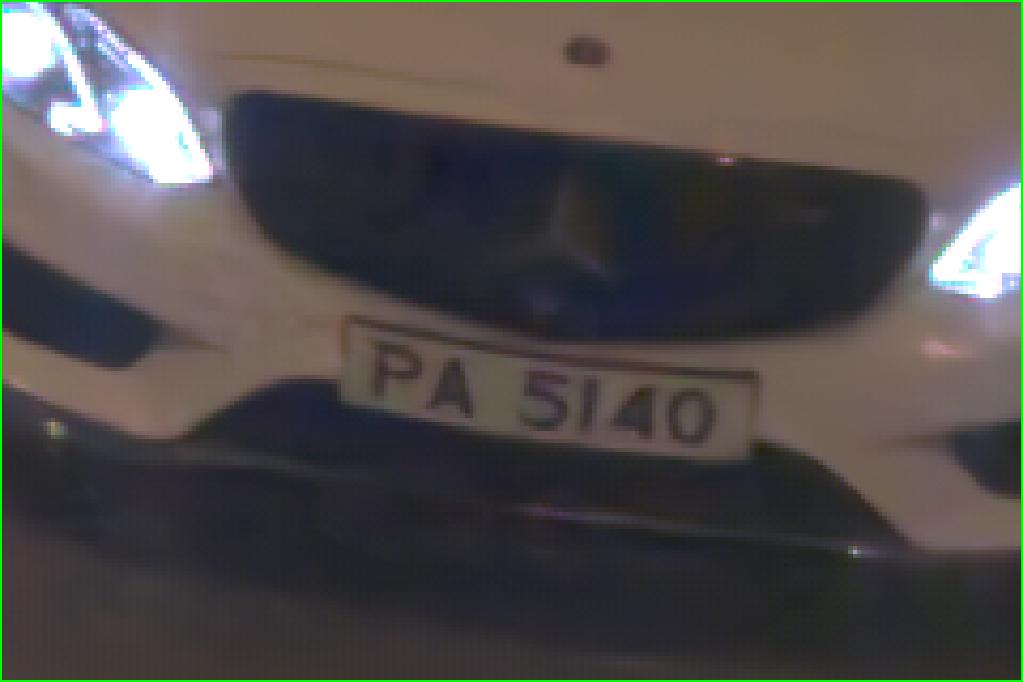}
    {\small RviDeNet+DMN}
    \end{minipage} %\hspace{0.1cm}
    \begin{minipage}[b]{0.24\textwidth}
    \centering
    \includegraphics[width=1\textwidth]{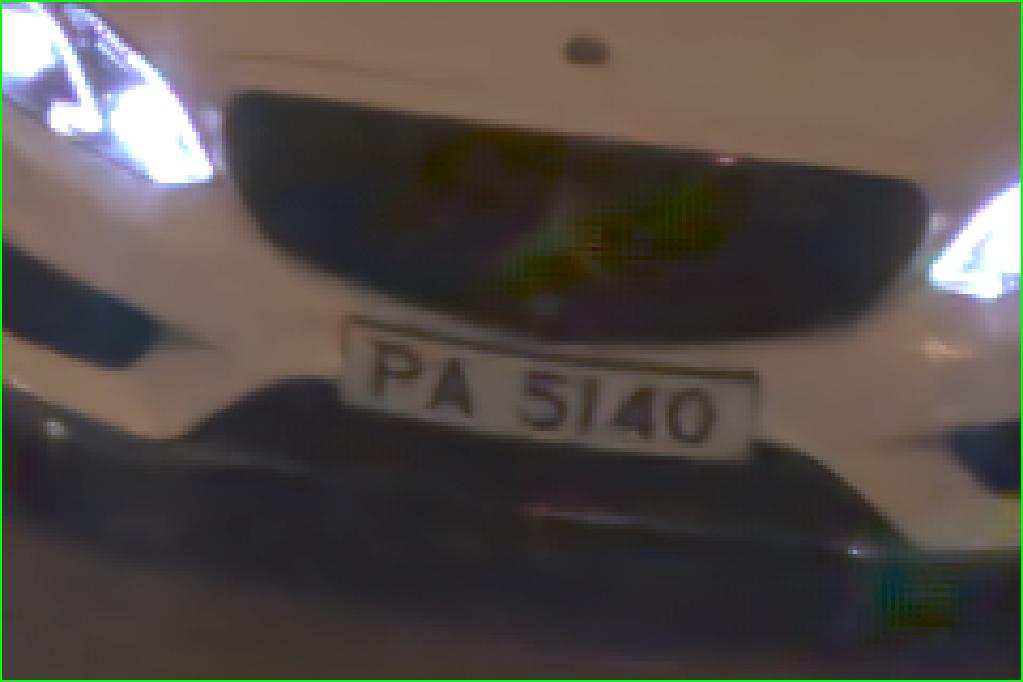}
    {\small RviDeNet*}
    \end{minipage} 
    \begin{minipage}[b]{0.24\textwidth}
    \centering
    \includegraphics[width=1\textwidth]{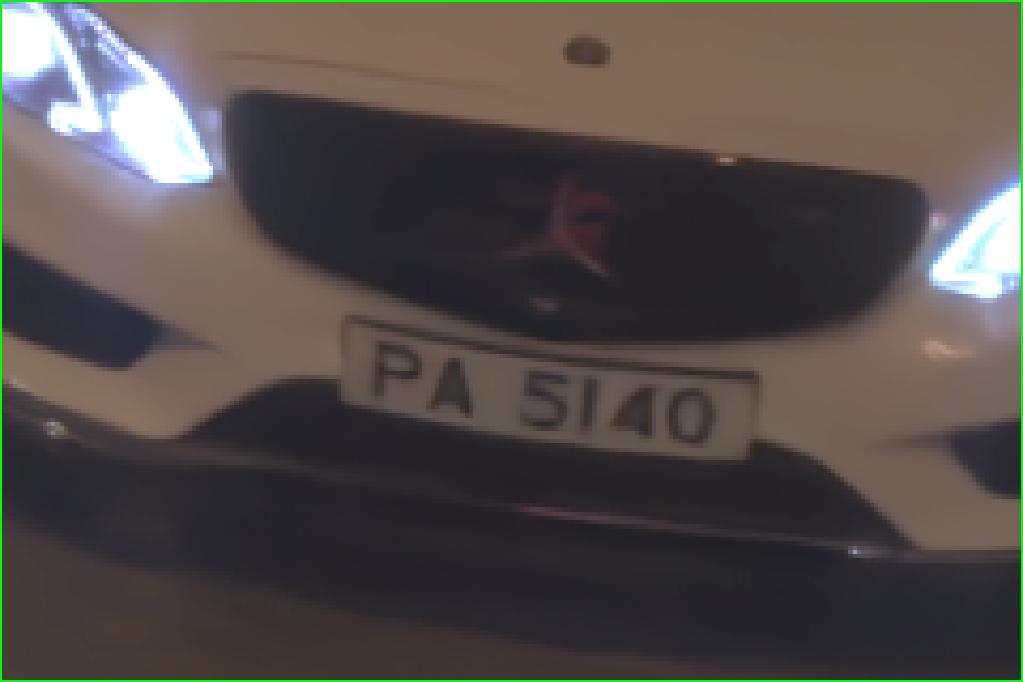}
    {\small Ours}
    \end{minipage} 
\end{minipage}%\hspace{0.1cm}
\vspace{0.15cm}

\caption{JDD results of different methods on burst images captured using iPhone X (first image), Pixel 2 (second image) and iPhone 7 (third image).}%\vspace{-0.4cm}
\label{realCaptured}
\end{figure*}

\subsection{Results on Synthetic Data}
\label{ExpSim}
Two video datasets, \emph{i.e.}, Vid4~\cite{liu2013bayesian} and REDS4~\cite{wang2019edvr}, are adopted in the experiments. Vid4 is widely used as the test set in the study of video super-resolution, and the video clips (resolution: $480\times720$) in Vid4 have small motion. The videos in REDS4 dataset have better quality and resolution (720p) but with bigger motions. Video clips are firstly converted to raw space using the unprocessing operator $\Gamma^{-1}(\cdot)$ introduced in Sec~\ref{implementdetails}. 
%Refer to \cite{brooks2019unprocessing}, we set the \emph{AWB gains} and \emph{xyz2cam} color matrix in $\Gamma^{-1}(\cdot)$ as [2.0,1.0,1.7] and [1.0234, -0.2969, -0.2266; -0.5625, 1.6328, -0.0469; -0.0703, 0.2188, 0.6406] for testing. 
Then the noise is added to the raw images by using Eq.~\ref{noisemodel}. Tables~\ref{synthVid4} and \ref{synthREDS4} list the PSNR/SSIM results of different algorithms under different noise levels. Following~\cite{mildenhall2018burst}, both PSNR and SSIM are computed after gamma correction to better reflect perceptual quality. One can see that the proposed GCP-Net achieves the best PSNR/SSIM measures. Visual comparisons on Vid4 and REDS4 are presented in Fig.~\ref{figVid} and Fig.~\ref{figRED}, respectively.

From Tables~\ref{synthVid4} and \ref{synthREDS4}, we can see that the performance of JDD-S methods FlexISP and ADMM is generally far below the learning based multi-frame algorithms. However, the multi-frame based VBM3D+DMN fails to compete with the single frame based ADMM, especially on the REDS4 dataset. This is mainly because VBM3D+DMN assumes AWGN and it cannot handle the large-motion in the videos of REDS4. The CNN-based methods KPN+DMN, EDVR+DMN and RviDeNet+DMN achieve much better performance than JDD-S methods because they can learn to handle the misalignments between adjacent frames and exploit the temporal redundancy for denoising. Nonetheless, one can still see that these methods generate noticeable color artifacts and zippers (see Figs.~\ref{figVid}(b)(c)(e) and Figs.~\ref{figRED}(b)(c)(e)) near edges and complex textures. This is mainly because they perform burst denoising and color demosaicking separately without considering the correlations between the two tasks. For JDD-B algorithms, \emph{i.e.}, EDVR* and RviDeNet*, their restoration results contain less zippering and color artifacts compared with EDVR+DMN and RviDeNet+DMN, which proves the effectiveness of jointly handling denoising and demosaicking task. However, the results of EDVR* and RviDeNet* suffer from the over-smoothing problem (see Figs.~\ref{figRED}(d)(f)). Benefiting from the GCP guidance, our GCP-Net performs the best in the JDD task, making good balance between noise removal and structure preservation (see Fig.~\ref{figRED}(g)).

\subsection{Results on Real-world Data}
\label{realexp}
We also compare different JDD algorithms using real-world burst raw images captured by several smartphone cameras, including iPhone 7, iPhone X and Pixel 2. Since there is no ground-truth for the collected images, we can only provide qualitative comparison. The restored images are converted to sRGB domain for visualization by using the ISP operations, including white balance, color transfer and gamma compression. The white balance parameters, color matrix and the noise level are collected from the camera metadata.

In Fig.~\ref{realCaptured}, we show the JDD results of several noisy images captured by the three smartphone cameras under normal lighting conditions as well as nighttime environments with a wide range of ISO values. Similar conclusions to the synthetic experiments can be made. KPN+DMN, EDVR+DMN and RviDeNet+DMN can remove noise but will produce zippering artifacts and smooth the image details. By jointly performing denoising and demosaicking, EDVR* and RviDeNet* can reduce the zippering effect but still produce over-smoothed reconstruction. In the zoom-in patch of the second image in Fig.~\ref{realCaptured}, we can also see that RviDeNet* generates artifacts in the high noisy area. Our proposed GCP-Net can effectively remove noise while retaining fine textures. It is also free of the moire pattern. In the last row of Fig.~\ref{realCaptured}, we show the JDD results on night-shot burst images with large motion. We can see that the restoration result of KPN+DMN contains much ghosting artifact, which is mainly caused by the large motion of object (\emph{i.e.}, car) in the burst images. RviDeNet* also generates some motion induced artifacts around the moving objects. The proposed GCP-Net can work stably under such large-motion scenes. This is because we utilize the pyramid offset estimation and the offset is estimated on GCP features which dilute the impact of noise.

\subsection{Model Size}
Table~\ref{params} lists the number of model parameters and the number of floating point operations (FLOPs) of our GCP-Net and the comparison methods, \emph{i.e.}, DMN, KPN, EDVR and RviDeNet. The FLOPs are calculated on 5 frames of size $128\times 128$. Because of the GCP branch and some attention modules, GCP-Net has more parameters than DMN, KPN and EDVR, but its model is much smaller than RviDeNet. Though the number of parameter and FLOPs of GCP-Net are 2 times and 1.5 times that of EDVR, it achieves significantly better JDD results than EDVR ($\sim$3.5dB on Vid4 and $\sim$2dB on REDS4). Since RviDeNet utilizes a pre-denoising network and non-local modules, it has very high computational cost but achieves lower JDD performance than GCP-Net. Overall, GCP-Net achieves a good balance between JDD effectiveness and efficiency. 

\begin{table}[!tbp]
\setlength{\abovecaptionskip}{0.2cm}
\footnotesize
%\scriptsize
\centering
%\smallskip
\caption{Comparison of different CNN models used in comparison in terms of the number of parameters and FLOPs on 5 frames of size $128 \times 128$}
\label{params}
%\begin{tabularx}{\linewidth} {c c c c c}
%\begin{tabularx} {\linewidth} { @{}  p{1.0cm}  X X X  X X @{}}
\begin{tabularx} {\linewidth} {P{1.2cm} P{1.0cm} P{1.0cm} P{1.0cm} P{1.0cm} P{1.2cm}}
\toprule
  & DMN & KPN & EDVR & RviDeNet & GCP-Net \\
\midrule
\#. Params & 0.56M & 5.58M & 6.28M & 57.98M & 13.79M \\
FLOPs & 2.2G & 4.2G & 55.0G & 482.2G & 78.8G \\
\toprule

\end{tabularx} 
\end{table}

\section{Conclusion}
\label{con}
Most of the previous joint denoising and demosaicking (JDD) methods worked on a single color filter array (CFA) image. In this paper, we proposed an effective network, namely GCP-Net, for JDD on real-world burst images (JDD-B). Our method took the advantages of green channel prior (GCP), which referred to the fact that the green channel of CFA raw images usually had higher quality and sampling rate than the red and blue channels. The GCP features were used to guide the intra-frame feature extraction, inter-frame fusion and upsampling process of the multi-frame JDD-B task. Our experiments on synthetic data and real-world data quantitatively and qualitatively demonstrated that GCP-Net achieved superior performance to existing state-of-the-art algorithms of JDD. It can remove the heavy noise from images captured in low-light condition, preserve the textures and details without generating much visual artifact.

%\appendices

%\section{}

%\section*{Acknowledgment}


\ifCLASSOPTIONcaptionsoff
  \newpage
\fi

\bibliographystyle{IEEEtran}
\bibliography{egbib}

\end{document}